\documentclass{mn2e}
\usepackage{times,graphicx,amssymb,color}
\def\arcsec{\hbox{$^{\prime\prime}$}}
\title[HD 40307 g]{A dynamical study on the habitability of terrestrial exoplanets II: The super Earth HD 40307 g}
\author[Brasser, Ida and Kokubo]{R. Brasser$^1$, S. Ida$^2$ and E. Kokubo$^3$\\
$^1$ Institute for Astronomy and Astrophysics and Theoretical Institute for Advanced Research in Astrophysics, Academia Sinica,
Taipei 10617, Taiwan\\ 
$^2$ Earth-Life Science Institute, Tokyo Institute of Technology, $\bar{O}$okayama, Meguro district, Tokyo 152-8551, Japan\\ 
$^3$ Division of Theoretical Astronomy, National Astronomical Observatory of Japan, Osawa, Mitaka district, Tokyo 181-8588, Japan\\ 
}
\begin{document}
\maketitle
\begin{abstract}
HARPS and {\it Kepler} results indicate that half of solar-type stars host planets with periods $P<100$~d and masses
$M<30$~$M_{\oplus}$. These super Earth systems are compact and dynamically cold. Here we investigate the stability of the super
Earth system around the K-dwarf HD40307. It could host up to six planets, with one in the habitable zone. We analyse the system's
stability using numerical simulations from initial conditions within the observational uncertainties. The most stable solution deviates
3.1$\sigma$ from the published value, with planets e and f not in resonance and planets b and c apsidally aligned. We study
the habitability of the outer planet through the yearly-averaged insolation and black-body temperature at the pole. Both undergo
large variations because of its high eccentricity and are much more intense than on Earth. The insolation variations are precession
dominated with periods of 40~kyr and 102~kyr for precession and obliquity if the rotation period is 3~d. A rotation period of about
1.5~d could cause extreme obliquity variations because of capture in a Cassini state. For faster rotation rates the periods converge to
10~kyr and 20~kyr. The large uncertainty in the precession period does not change the overall outcome.
\end{abstract}

\section{Introduction}
Since the discovery of the first extrasolar planet in 1995 (Mayor \& Queloz, 1995) there has been a surge in research in
planetary science and in the detection of new planets, with the HARPS survey (Mayor et al., 2003) and NASA's {\it Kepler} mission
leading the field. The high number of detected and candidate planets allows for statistical studies and several trends have emerged,
which are shared among both the HARPS and Kepler data (Figuera et al., 2012). Some of these include:

\begin{itemize}
\item Approximately half of all solar-type stars contain planets with a projected mass $m_p \sin I <30$~Earth masses ($M_\oplus$)
(Borucki et al., 2011; Mayor et al., 2011; Chiang \& Laughlin, 2013), where $m_p$ is the planet's mass, and $I$ is the angle
between the planet's orbit and the observer.
\item Planets with a short orbital period tend to be of low ($<30$~$M_\oplus$) mass (Mayor et al., 2011; Batalha et al., 2013). Most
of these have radii $R \in [1,4]$~Earth radii ($R_\oplus$) and masses between Earth's and Neptune's. These planets are often referred
to as super Earths. In contrast, hot Jupiters are rare (Mayor et al., 2011).
\item The number of planets increases with decreasing mass and/or radius (Howard et al., 2010; Howard et al., 2012) and approximately
23\% of stars have Earth-like close-in planets with periods $P<50$~days (d).
\item Approximately 73\% of low-mass planets with periods shorter than 100~d reside in multiple systems (Mayor et al.,
2011; Fang \& Margot, 2012), with only 26\% of these multiples containing a gas giant (Mayor et al., 2011). This suggests that
super Earths form in clusters close to the parent star and are isolated from potential giant planets in the system.
\item Systems of multiple super Earths on short periods tend to be compact (Fang \& Margot, 2012; Chiang \& Laughlin, 2013) and have
low ($<3^\circ$) mutual inclinations (Fang \& Margot, 2012; Tremaine \& Dong, 2012) and most likely also low ($<0.2$) eccentricities
(Mayor et al., 2011; Wu \& Lithwick, 2013).
\item The period distribution is more or less random with some excesses just slight of the 3:2 and 2:1 mean motion resonances (Fabrycky
et al., 2012). The near-resonance of some pairs has been attributed to tidal decay (Batygin \& Morbidelli, 2013; Lithwick \& Wu,
2012), { though Petrovitch et al. (2013) proposed an alternative scenario based on planet growth}.
\item Although still actively debated, the period distribution of exoplanets with short ($P<200$~d) periods suggests an
in-situ formation scenario (Raymond et al., 2008; Hansen \& Murray, 2012; Chiang \& Laughlin, 2013) rather than formation farther out
followed by migration (Lopez et al., 2012; Kley \& Nelson, 2012; Rein, 2012). An intermediate scenario in which planetary embryos
migrate inwards followed by a giant impact stage (Ida \& Lin, 2010) may also work.
\item The mutual spacing of most of these super Earths is between 5 to 30 Hill radii (Lissauer et al., 2011; Fabrycky et al., 2012),
which encompasses the spacing between the giant planets (12) and terrestrial planets (40). However, their proximity to the star
requires their orbits to be dynamically cold to prevent orbit crossing.
\item The Kepler catalogue contains a few confirmed super Earth planets in the habitable zone of their parent stars, { with a
further 20 candidates} (Batalha et al., 2013). The habitable zone (HZ) is the region where radiation received by the planet from the
star is enough for it to sustain liquid water under sufficient atmospheric pressure (Kasting et al., 1993; Kaltenegger \& Sasselov,
2011).
\end{itemize}
Thus, it seems the super Earth population resembles the regular satellite populations of the giant planets: both show a typical mass
ratio of $m_p/M_* \sim 10^{-4.5}$ and regularly spaced, dynamically cold orbits. Some are far enough out to be in the habitable
zone.\\
The regularity of the orbits and tight spacing provide a formidable challenge to theorists of planet formation and dynamicists alike.
If the best determined orbits are not entirely circular determining whether or not these systems are dynamically stable and fall within
the observational uncertainties is challenging. The aim of this study is to analyse the stability of one such compact super Earth
system: HD 40307. This system is interesting because one planet, HD 40307 g, may be in the habitable zone. Therefore we also study
how the dynamics of the whole system affects the long-term habitability of planet g.\\
The term `habitable' encompasses many things, however, thus we only focus on the long-term variations in the insolation caused by the
dynamics of the whole system and the solid body response of planet g. On Earth, in addition to stellar activity, geological activity
and temperature regulation through the carbonate-silicate cycle (Williams \& Kasting, 1997), the long-term climate is driven
externally by the Milankovi\'{c} cycles (Milankovi\'{c}, 1941). Earth's orbit is perturbed by other planets, causing quasi-periodic
variations in eccentricity and inclination on a time scale of 100~kyr. The Earth's obliquity is also affected and oscillates on a time
scale of 41~kyr (e.g. Laskar et al., 1993). The combined effect of the variations in eccentricity, obliquity and precession angle are
the Milankovi\'{c} cycles. Those periods of the Milankovi\'{c} cycles are much shorter than the relaxation time of the
carbonate-silicate feedback mechanism.\\
The perturbations of other planets cause these changes and so influence the insolation accordingly, driving the ice ages on the Earth
(Imbrie \& Imbrie, 1980). Small variations in eccentricity and obliquity most likely yield stable and favourable conditions for
habitability (Atobe et al., 2004; Brasser et al., 2013).\\
We want to know what are the dynamical properties of close super Earth systems and how dynamics affects the habitability of potentially
habitable planets. This paper is a proof of concept on how to determine the dynamical stability of a compact super Earth system, and
how the long-term insolation variation of any planets in the habitable zone depends on the dynamics of the whole system. Other
effects, such as a general circulation model (GCM) of the planet, and affects of atmospheric heat transport and buffering, ice-albedo
feedback, carbon dioxide cloud formation, are not a part of this study but will be part of future projects.\\
This paper is organised as follows. The next section contains an overview of the HD 40307 planetary system. In Section 3 we describe
our numerical methods. In Section 4 we summarise the theory of how orbital perturbations affect the obliquity and the
black-body equilibrium temperature of the planet. This is followed by our results in Section 5. Section 6 focuses on the long-term
climate cycles and Section 7 is reserved for a discussion. We present a summary and conclusions in Section 8.

\section{System overview}
\begin{table*}
\center
\caption{Orbital element solution of the HD 40307 system from Tuomi et al. (2013) with updated mean longitudes. We list the means and
standard deviations, whenever available, taken both from their tables and figures. When two values are listed in the square brackets
these correspond to minimum and maximum values respectively. Unlike Tuomi et al. (2013) we did not consider variations in the mass of
the star and thus our deviations in the semi-major axis are smaller than theirs. The equilibrium temperatures are computed by us. The
value in brackets is the orbit-averaged value where the nominal eccentricity was used. We have taken the albedos to be 0.3, the
stellar surface temperature was $T_* = 4700$~K and the stellar radius was 0.91$R_{\odot}$ and its luminosity is $L=0.23L_\odot$.}
\label{orbits}
\begin{tabular}{lccc}
\hline \hline
Parameter ($\mu$,[$\sigma$]) & HD 40307 b & HD 40307 c & HD 40307 d \\
\hline
$P$ [days] & 4.3123 [3.9167$\times 10^{-4}$] & 9.6183 [1.6825$\times 10^{-3}$] & 20.4321 [7.5606$\times 10^{-3}$] \\
$e$ & 0.1964 [0.054187] & 0.05777 [0.03609] & 0.06728 [0.03514] \\
$\omega$ [$^\circ$] & 194.8 [137.5, 240.6] & 234.9 [-] & 17.2 [-] \\
$\lambda$ [$^\circ$]& 150.0 [-] & 5.0 [-] & 353.0 [-] \\
$m_{p} \sin I$ [M$_{\oplus}$] & 4.0 [3.3, 4.8] & 6.6 [5.6, 7.7] & 9.5 [8.0, 11.2] \\
$a$ [AU] & 0.047524 [3$\times 10^{-6}$] & 0.08113 [2$\times 10^{-5}$] & 0.13407 [3 $\times 10^{-5}$] \\
$\langle T_{\rm eq} \rangle$ [K] & 910 [1217] & 696 [792] & 541 [625] \\
\hline
& HD 40307 e & HD 40307 f & HD 40307 g \\
\hline
$P$ [days] & 34.6279 [6.9005$\times 10^{-2}$] & 51.7768 [0.1721] & 196.343 [2.1817] \\
$e$ & 0.1201 [0.06653] & 0.06175 [0.04534] & 0.2205 [0.1113] \\
$\omega$ [$^\circ$] & 303.7 [-] & 355.2 [-] & 91.7 [-] \\
$\lambda$ [$^\circ$]& 62.0 [-] & 308.0 [-] & 127.0 [-] \\
$m_{p} \sin I$ [M$_{\oplus}$] & 3.5 [2.1,4.9] & 5.2 [3.6, 6.7] & 7.1 [4.5, 9.7] \\
$a$ [AU] & 0.1906 [2 $\times 10^{-4}$] & 0.2491 [6 $\times 10^{-4}$] & 0.6069 [9.6$\times 10^{-4}$] \\
$\langle T_{\rm eq} \rangle$ [K] & 455 [558] & 398 [423] & 255 [349] \\
\hline
$\chi^2=2.5$ & JD 2450000 & &\\
\hline
\hline
\end{tabular}
\end{table*}
\begin{figure}
\resizebox{\hsize}{!}{\includegraphics{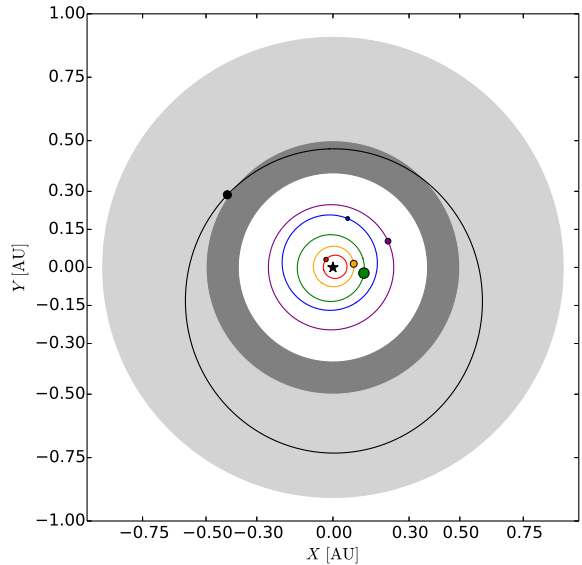}}
\caption{A top-down view of the orbits of the planets orbiting HD 40307 drawn to scale. The positive $x$-axis marks zero longitude.
The shaded area marks the habitable zone. The positions of the planets are taken from Table~\ref{orbits}. The planets' marker sizes are
proportional to their masses.}
\label{orb}
\end{figure}
From high resolution spectroscopic data with the HARPS instrument at ESO's 3.6 m telescope at La Silla, the K-dwarf star HD 40307
was found to have three planetary companions on near-circular orbits with periods ranging from 4~d to 20~d and masses from 4 to 10
Earth masses on dynamically stable orbits (Mayor et al., 2009). A subsequent re-analysis of the data with the HARPS-TERRA algorithm 
(Anglada-Escude \& Butler, 2012) and the inclusion of further observations led Tuomi et al. (2013) to conclude that HD 40307 may be
surrounded by up to six super Earth planets, with the outermost one being in the habitable zone. Figure~\ref{orb} shows a top-down
view of the orbits of the HD 40307 system, using the orbital parameters of Tuomi et al. (2013). The star is surrounded by a compact
inner five-planet system of super Earths and a last planet farther out in the habitable zone on a fairly eccentric orbit.
The habitable zone is indicated by light grey shading. In this paper we analyse the dynamical stability of the system and the
obliquity and insolation forcing on planet g. In order to do so we need a dynamically stable solution that best matches the HARPS data.
However, when fitting the orbital solution of Tuomi et al. (2013) to the radial velocity data, we computed a reduced $\chi^2 \gg 10$,
indicating a poor fit and casting doubt on the published solution. We attempted to reduce it and still be consistent with the radial
velocity data.\\ 
The radial velocity at any epoch is given by
\begin{equation}
V_r = V_0 + \sum_{k=1}^6\frac{m_k}{M_*} \frac{2\pi a_k}{P_k}\frac{\sin I}{\sqrt{1-e_k^2}}[\cos(\upsilon_k+\omega_k)+e_k\cos \omega_k],
\label{rv}
\end{equation}
where $\upsilon_k$ is the planet's true anomaly, $\omega_k$ is its argument of periastron, $a_k$ is the semi-major axis, $P_k$ is the
orbital period, $m_k$ is the planetary mass, $M_*$ is the stellar mass, $I$ is the inclination of the orbital plane with respect to
the viewer and $V_0$ is a non-zero drift velocity. Apart from the periods, the greatest variation in $V_r$ is caused by changing
$\upsilon_k+\omega_r$, and thus we used a simple grid search, varying the initial mean longitude of each planet, until we found the
minimum $\chi^2 = 2.5$. The orbital elements corresponding to this minimum are given in Table~\ref{orbits}. We list the mean values
($\mu$) and standard deviations ($\sigma$), whenever available, of the following quantities: the orbital period, $P$, eccentricity,
$e$, argument of periastron, $\omega$, mean longitude at the epoch, $\lambda$, inferred planetary masses $m_p \sin I$, and semi-major
axes, $a$. The semi-major axes are computed directly from the period and the standard deviations assume a Gaussian fit (Tuomi et al.,
2013).\\
To obtain an overview of the system, we use some of the quantities introduced by Chambers (2001): the Angular Momentum
Deficit (AMD) (Laskar, 1997), fraction of total mass in the most massive planet ($S_m$), a spacing parameter ($S_s$) that scales as
the planet to star mass ratio $\mu^{1/4}$ rather than the Hill relation of $\mu^{1/3}$ (Chambers et al., 1996), and a concentration
parameter ($S_c$) which measures how much mass is concentrated in a narrow annulus. We supplement this with listing the average spacing
in Hill radii ($S_H$). We calculated these values for the entire system, the inner five planets, the terrestrial
planets and the giant planets of our Solar System in Table~\ref{chamb}.\\
\begin{table}
\label{chamb}
\begin{tabular}{c|ccccc}
  System & AMD & $S_m$ & $S_s$ & $S_c$ & $S_H$ \\ \hline \\
  HD 40307 & 9.9 $\times 10^{-3}$& 0.265 & 21.7 & 8.29 & 19.6\\
  HD 40307 b-f & 2.1 $\times 10^{-3}$& 0.330 & 19.1 & 18.7 & 16.3\\
  MVEM & 1.8 $\times 10^{-3}$ & 0.509 & 37.7 & 89.9 & 43.2 \\
  JSUN & 1.3 $\times 10^{-3}$ & 0.715 & 11.5 & 27.0 & 12.0
 \end{tabular}
\caption{Chambers' (2001) stability quantities for the entire HD 40307 system, the inner five planets, and for comparison also the
terrestrial planets and giant planets of our Solar System.}
\end{table}
The AMD of HD 40307 is about a factor five higher than either the terrestrial planets and giant planets of the Solar System. This is
dominated by the high eccentricity of planets b and g. Systems with higher AMD are possibly more chaotic (Laskar, 1997) and have
more opportunity to exchange it among the planets, resulting in larger eccentricity oscillations. However, due to the large spacing
between planet g and the innermost planets, secular interaction between these two groups will be minimal.\\
Given that all planets have comparable mass, $S_m$ is lower than in the Solar System. The spacing parameter is comparable to that of
the giant planets. However, the mass concentration parameter, $S_c$ is the opposite: it is only of the order of 10, while in the Solar
System's it ranges from 27 to 90. The low value of $S_c$ indicates that all planets have a comparable mass and that the total mass is
spread over a wide annulus rather than concentrated in a narrow one composed of two heavy planets.\\
\begin{table}
\begin{tabular}{c|c}
  Period ratio & Value (median; standard deviation)\\ \hline \\
  $P_{\rm c}/P_{\rm b}$ & 2.2304 [5$\times 10^{-4}$] \\
  $P_{\rm d}/P_{\rm c}$ & 2.1243 [4$\times 10^{-4}$] \\
  $P_{\rm e}/P_{\rm d}$ & 1.6948 [3 $\times 10^{-3}$] \\
  $P_{\rm f}/P_{\rm e}$ & 1.4952 [6 $\times 10^{-3}$]
\end{tabular}
\caption{Period ratios between some inner planet pairs. Only the planets e and f appear to be in a resonance.}
\label{prat}
\end{table}
We also inspect the period ratios between the planets for mean-motion resonances. These are listed in Table~\ref{prat}. The planets e
and f appear close to a 3:2 mean motion resonance. Tuomi et al. (2013) concluded that the planets e and f need to be in a 3:2 resonance
to be dynamically stable. The other planets do not appear to be in resonance, but Papaloizou \& Terquem (2010) state that they are
close enough for their dynamics to be affected as if they were in resonance.\\
Another feature warranting discussion is the relatively high eccentricity of planet b. One would expect that its proximity to the
star would have fully damped its eccentricity due to tidal effects. However, as shown by Lovis et al. (2011) and subsequently
detailed by Batygin \& Laughin (2011), tidal eccentricity damping is more complicated than simply reducing the eccentricity to zero.
Tidal friction damps the free eccentricity mode of planet b (mode 1) the fastest; the damping of the other (forced) modes takes much
longer. Thus, it is most likely that the eccentricity of planet b is caused entirely by forcing from the other planets, mostly from
planet c. Batygin \& Laughin (2011) demonstrate that in a two-planet system the complete damping of the mode corresponding to the
innermost planet leads to apsidal alignment or anti-alignment of the two planets with the eccentricities in both planets having a fixed
ratio; the actual configuration depends on the relative magnitudes of the two eigenfrequencies.\\
We can predict the configuration of the innermost two planets from theory. The tidal damping time scale of the eccentricity of
the innermost planet can be obtained from Ferraz-Mello's (2013) equilibrium tidal theory. For a quasi-stationary rotation in which the
semi-diurnal forcing frequency, $f = 2\nu - 2n$, is low and $O(e^2)$, we have
\begin{equation}
\tau_{\rm eb} = \frac{32}{1125}\frac{m_{\rm{b}}R_{\rm b}^2}{\mathcal{C}}\frac{m_{\rm b}}{M_*}\Bigl(\frac{a}{R_{\rm
b}}\Bigr)^5\Bigl(1+\frac{\gamma^2}{n^2}\Bigr)\Bigl(\frac{\gamma}{1\,\rm{rad}~\rm{yr}^{-1}}\Bigr)^{-1},
\end{equation}
where $\nu$ is the planet's rotation frequency, $n=\sqrt{GM_*/a^3}$ the mean motion, $\mathcal{C}$ the largest moment of
inertia, $M_*$ the stellar mass, $a$ the semi-major axis, $m_{\rm{b}}$ and $R_{\rm{b}}$ are the mass and radius of planet b and
$\gamma$ is the tidal relaxation frequency. If we assume that $\gamma \sim 1.8 \times 10^{-7}$~Hz ($\sim 35.7$~rad~yr$^{-1}$) as for
solid Earth and that the planet follows the mass-radius relation $R_{\rm b}/R_{\oplus} \sim (m_{\rm b}/m_\oplus)^{0.274}$ (Valencia et
al., 2006; Sotin et al., 2007), which appears consistent with observations of super Earth planets (Gaidos et al., 2012), we have
$\tau_{\rm eb} \sim 10$~Myr. Thus the eccentricity associated with mode 1 should be fully damped in approximately $\sim 50$~Myr,
much shorter than the age of the system. This value is uncertain by a factor of a few, with the largest uncertainty coming from the
estimate of the planetary radius. This short eccentricity damping time scale is somewhat problematic because over the course of
several Gyr the corresponding tidal decay of the semi-major axis is $\sim 0.02$~AU assuming $e \sim 0.1$. This poses constraints on the
formation and past evolution of the system that are beyond the scope of this paper.\\
The eccentricity mode corresponding to planet c will decay on a time scale of at least $\tau_{\rm ec} \sim 200$~Myr, so that this mode
may have also decayed since the formation of the system. However, given the uncertainty in $\gamma$ it is not evident that the mode
corresponding to planet c has also decayed away. We have used $\gamma$ for the solid Earth because the size of the planets suggest they
may be partially molten inside. But when setting $\gamma$ equal to the values for Mars and the Moon, which are practically solid
bodies, the decay times become one to two orders of magnitude longer: still short enough to damp the eigenmode corresponding to
planet b but not that of planet c.\\
\begin{table}
\begin{tabular}{cccccc}
$g_1$ [$\arcsec$~yr$^{-1}$] &$g_2$ & $g_3$ & $g_4$ & $g_5$ &$g_6$\\
\hline \\
1834 & 1018 & 232 & 1541 & 561 & 16.7 \\ \\
$s_1$ [$\arcsec$~yr$^{-1}$] &$s_2$ & & $s_4$ & $s_5$ &$s_6$\\ \hline \\
-2036 & -1002 & 0.0 & -1615 & -488  & -25.5
\end{tabular}
\caption{Eigenfrequencies $g_{\rm n}$ and $s_{\rm n}$of the HD 40307 planetary system obtained using the Laplace-Lagrange AMD theory
of Agnor \& Lin (2012). The frequencies depend on the original configuration of the orbits i.e. the frequencies increase if planets are
more eccentric and thus spend more time in each other's presence, or if orbits are nearly anti-aligned versus being aligned.}
\label{eigenval}
\end{table}

\begin{table}
\begin{tabular}{c|cccccc}
&& &$S_{i,j}$& &&\\
$j$/$i$ & 1 & 2 & 3 & 4 & 5 & 6 \\
\hline \\
1 & 0.5311 & -0.6678 & 0.1451 & 0.3469 & -0.3641 & 0.0024 \\
2 & -0.7126 & -0.3643 & 0.2669 & -0.2282 & -0.4860 & 0.0050 \\
3 & 0.3699 & 0.5103 & 0.5021 & -0.3158 & -0.5008 & 0.0133 \\
4 & -0.2659 & 0.3019 & 0.4066  & 0.8202 & 0.0018 & 0.0157 \\ 
5 & 0.0507 & -0.2641 & 0.6997 & -0.2348 & 0.6184 & 0.0242 \\
6 & -0.00003 & 0.00054 & -0.02903 & 0.0001 & -0.0060 & 0.9996 \\ \hline \\
$C^2_j$ & 0.01701 &  0.1285 & 0.03871 & 0.03999 & 0.01071 & 0.7650\\ 
$\beta_j$ & 118.6$^\circ$ & 10.1$^\circ$ & 320.7$^\circ$ & 253.9$^\circ$ & 23.1$^\circ$ & 91.4$^\circ$
\end{tabular}
\caption{Eccentricity eigenvectors (columns) of the Laplace-Lagrange AMD solution (Agnor \& Lin, 2012) for the nominal orbital
parameters listed in Table~\ref{orbits}. The rows depict the forcing of the planets on each other. The integration constants (amplitude
and phases) are listed separately in the bottom.}
\label{eigenvec}
\end{table}
Unfortunately planets e and f could be in a 3:2 resonance so that the often-used Laplace-Lagrange theory cannot accurately predict
their apsidal precession frequencies. Nevertheless, we applied the theory to the nominal orbital elements of the system and computed
the eigenfrequencies.  We used the modified AMD Laplace-Lagrange theory from Agnor \& Lin (2012), which gives a better overview of
the amount of AMD in each eigenmode. We included the effect of General Relativity following Batygin \& Laughin (2011). We did not
include the influence of tides or additional effects arising from the non-spherical shape of the planets because these are all
insignificant compared to General Relativity (Batygin \& Laughlin, 2011). We listed the eccentricity ($g_{\rm i}$) and inclination
($s_{\rm i}$) eigenfrequencies in Table~\ref{eigenval}. Unlike in the Solar System, the strong coupling between the inner five planets
makes identification of any frequency with a particular planet rather difficult and the identification that is listed is a result
of numerical simulations. The corresponding eccentricity eigenvectors, with components $S_{i,j}$, are listed as the columns of
Table~\ref{eigenvec}, with the normalised integration constants $C^2_j$/AMD (Agnor \& Lin, 2012), $\beta_j$ (amplitudes and phases)
given separately below. The rows identify the strength of the forcing on a particular planet caused by the other planets. The total
entropy is $Z \equiv -\sum_{j=1}^{2N-1} C^2_j \ln C^2_j$ (Wu \& Lithwick, 2011), which evaluates to $\exp(Z)\sim2.3$ out of a maximum
of 11. The minimum value of $Z$ is 0, corresponding to the AMD being confined to one single eigenmode, while $Z$ approaches $\ln 11$ at
equipartition. The fairly low value of $Z$ suggests that the AMD is confined to two eigenmodes (modes 2 and 6) and the system has
undergone some AMD diffusion in the past. Since we consider a planar system only, the value of $Z$ is most likely to be higher. For the
inner five planets only $\exp(Z) \sim 3.5$ and three eigenmodes are activated (modes 2, 3 and 4). From the top row of
Table~\ref{eigenvec} one may see that the eccentricity of planet b contains strong forcing from all eigenmodes, but mostly from mode 2.
Thus it is not surprising that planet b has a fairly high eccentricity. Assuming that this eccentricity is mostly a forced component
from mode 2, and numerical simulations demonstrate that the eccentricity of planet c is aligned with eigenmode 2, then the planets b
and c are likely to reside in apsidal alignment. Indeed, the components $S_{2,1}$ and $S_{2,2}$ carry the same sign and $g_1>g_2$ so
that the nominal solution of Table~\ref{orb} places planets b and c in apsidal alignment.\\
Note also that $C^2_2 \gg C^2_{1,3,4,5}$ which begs the question whether the excitement of eigenmode 2 is a remnant of formation, of
chaotic AMD diffusion or caused by a past 2:1 resonance passage with planet b as the latter spirals towards the star by tidal
interaction. If it is the latter this passage should have occurred recently because the eccentricity in planet c will decay away on a
time
scale of $O$(1~Gyr).

\section{Numerical methods}
\label{nm}
We analysed the stability of the HD 40307 system with a high number of numerical simulations. Our procedure was as follows.\\
First, we generated a new set of orbital parameters within the observational uncertainties, restricting ourselves to the deviations
in the orbital periods, eccentricities and arguments of periastron. We assumed the latter were evenly distributed from 0 to
360$^\circ$. We kept the mean longitude constant and equal to their nominal values.\\
We sampled the semi-major axis or eccentricity evenly between -3$\sigma$ and 3$\sigma$. This procedure was repeated a second time
for another orbital element resulting in an evenly-spaced 2-dimensional grid in $a-e$, $a-\omega$ or $e-\omega$. For each set of
simulations we only changed the orbital elements of a single planet, keeping the orbital elements of the others at their nominal
values. A typical grid contained 12\,000 entries.\\
Second, we integrated the system and analysed its stability in a manner similar to Correia et al. (2005). We simulated each
fictitious system for 40~kyr using the integrator SWIFT MVS (Levison \& Duncan, 1994), which is based on the fast and reliable
Wisdom-Holman method (Wisdom \& Holman, 1991). General relativity was included by adding the effect of a perturbing potential that
yields the correct periastron precession (Nobili \& Roxburgh, 1986), but does not reproduce the increased orbital frequency (Saha \&
Tremaine, 1994). This potential is
\begin{equation}
V_{\rm GR}=-3\Bigl(\frac{GM_*}{c}\Bigr)^2\frac{a}{r^3},
\end{equation}
where $G$ is the gravitational constant, $M_*$ is the stellar mass, $c$ is the speed of light and $r$ is the planet-star distance.
The time step was set to $3\times 10^{-4}$~yr or 2.6~h, approximately 40 steps for the innermost planet's orbit. Output was
generated every 40~yr. The simulation was stopped once a planet was farther than 1~AU from the star or once two planets entered each
other's Hill spheres. SWIFT MVS cannot handle close approaches between the planets when they are closer than their mutual Hill
spheres, and thus some simulations ended prematurely when such an encounter occurred.\\ 
Our simulation time was motivated by the following argument. When two planets are on planar crossing orbits, the probability of
the planets having an encounter within their Hill spheres is of the order of $(\frac{R_{\rm H}}{\pi a})^2$, where
$R_{\rm H}=a(m_{\rm p}/3M_*)^{1/3}$ is the size of the Hill sphere of each planet. For this system typically $m_{\rm p}/M_* \sim 2
\times 10^{-5}$ and thus the encounter probability is $P_{\rm enc} \sim 4 \times 10^{-5}$. Hence the time between encounters is $P_{\rm
enc}^{-1} \sim 20$~kyr and thus most unstable configurations should be detected within this time. We ran some test simulations for
20~kyr and 40~kyr and concluded that the latter resulted in more reliable stability maps. Thus we ran for 40~kyr.\\
We analysed the stability using a variant of Laskar's frequency analysis method (Laskar, 1993). The advantage of the frequency
analysis method is that it does not require long-term simulations and thus large regions of phase space can be tested in a reasonably
short time. Frequency analysis relies on computing a diffusion index, $D$, which is the fractional difference of the orbital frequency
of a planet averaged over two consecutive time intervals whose total time is the simulation time i.e. $T_1=T_2=T/2$. Therefore $D=\vert
n_1-n_2\vert$, where $n_1$ is the orbital frequency measured during the first interval, and $n_2$ is the orbital frequency during the
second interval. {However, while frequency analysis has been applied to study the stability of many systems it often does not
determine stability directly because the integrations are too short for the planets to have encounters and for the system to relax. As
a result, some long-term stable trajectories may be flagged as chaotic over short durations.}\\
We decided to approximate the orbital frequencies with the averaged mean motions over each interval {using Kepler's
third law} and follow the procedure of Marzari et al. (2003). We compute the average of the mean motion $\langle n_i \rangle$ over $N$
time intervals each of length $T_i=T/N$ and then compute the standard deviation, $\sigma_n$, over these $N$ intervals. The stability
index is then given by $D=\log(\sigma_ n/n_{\rm orig})$, where $n_{\rm orig}$ is the original mean motion that corresponds to the
published orbital period. We tested the procedure with five, ten and twenty intervals and concluded that twenty intervals resulted in
the best stability map. We did not perform any tests with more intervals.\\
{We realise that the interpretation of our method is not the same as with the frequency analysis. The latter measures diffusion in
the frequency space, while our method considers diffusion in action space through the relation between $n$ and $a$. Although for
regular motion the frequencies are constant and well defined, the actions such as the semi-major axis usually undergo small
oscillations so that in theory it is not adapted for diffusion measurements. However, in practise the average semi-major axis, being an
action of the Hamiltonian, is approximately constant along regular trajectories and its standard deviation could be a proxy for chaos
and thus for instability.} We repeat that here we are interested in determining the stability of the system rather than whether or not
it is chaotic. The stability is compromised by planetary encounters. The value of $D$ is directly related to the strength of a
planetary encounter, and very close encounters just outside the Hill sphere can lead to high values of $D$, much higher than the
deviations caused by approximating the mean motion with the orbital frequency. Thus our procedure is justified.\\ 
In addition to obtaining the index $D$ we also compute $D_{\rm max} = {\rm max}(D_{\rm j})$ i.e. the maximum value of all
six $D$-values of the planets. Low values ($D_{\rm max} \lesssim$ -3) correspond to a quasi-stable solution while high values ($D_{\rm
max} \gtrsim$ -2) are synonymous with unstable motion. When the simulation did not reach the final time because of a close encounter
we arbitrarily set $D_{\rm max}=0$.\\
We also compared the orbital configurations with the radial velocity data using the methods of Beaug\'{e} et al. (2012). For each pair
of elements in the stability map we calculated the radial velocity of HD 40307 at each epoch for which data was available. We
test the goodness of fit of the orbital elements to the radial velocity data using the reduced $\chi^2$. {The radial velocity data
used here is the same as that of Tuomi et al. (2013) and contains the same error sources.} However, as discussed in Beaug\'{e} et al.
(2012), care should be taken in blindly using $\chi^2$ because it assumes the errors are normally distributed, which
is often not the case. However, for want of a better descriptor, we employ it here with this caveat in mind.\\
Once one set of simulations was completed, we determined which initial values of orbital elements yielded the most stable
system that best agreed with the radial velocity data. We then updated the initial orbital elements of the planet whose orbital
elements were varied with those that gave the most stable configuration. We then used these updated elements for the next set of
simulations. This semi-iterative procedure ensures that each successive set of simulations has a high probability of converging to a
stable solution while still residing within the observational uncertainties.\\
The above steps are repeated until we have found a sizeable region of phase space where the system could be stable. We simulate the
most stable configuration for 1~Myr to determine its longer-term stability and use its output to determine the long-term climate
cycles of planet g below. All simulations were performed on the HTCondor pool at the Institute for Astronomy and Astrophysics, Academia
Sinica.\\

\section{Basic theory of how orbital deviations affect long-term climate trends}
\label{theory}
In this section we summarise how long-period orbital variations in a planet's orbit can affect the insolation on long time scales. We
shall discuss the case on Earth first since this is the best-studied example and then summarise how these effects could apply to
planet g.\\
It has been suggested that the long-term stability of the Earth's climate is related to the Milankovi\'{c} cycles (Milankovi\'{c},
1941), which cause insolation variations on long time scales due to changes in the Earth's eccentricity, obliquity and precession
angle. These variations manifest themselves the most at high latitudes. The onset and disappearance of ice ages appear correlated
to the insolation at summer solstice rather than the yearly-averaged insolation (Imbrie, 1982; Huybers, 2011). Each of these
three perturbations cause insolation forcing with specific periods of 100~kyr for the eccentricity, 41~kyr for the obliquity and
23~kyr for the precession. The recent Pleiostene ice ages record appears to be paced by both obliquity and precession forcing
(e.g. Huybers \& Wunsch, 2005; Huybers, 2011; Abe-Ouchi et al., 2013). To determine the insolation changes on planet g, we need to know
how its
obliquity and insolation change with time.\\
Introducing $\chi =\xi +\imath\eta= \sin \varepsilon\,\exp(\imath\psi)$, where $\varepsilon$ is the obliquity of the planet, $\psi$ is
the node of the planet's equator with its orbit (the precession angle) and $\imath$ is the imaginary unit, the equations of motion of a
planet's obliquity variations caused by perturbations from other planets on its orbit are (Neron de Surgy \& Laskar, 1997) 
\begin{eqnarray}
\dot{\xi} &=& A(t)\sqrt{1-\xi^2-\eta^2} -\eta[\alpha \sqrt{1-\xi^2-\eta^2} -2\Gamma(t)], \nonumber \\ 
\dot{\eta} &=& -B(t)\sqrt{1-\xi^2-\eta^2} +\xi[\alpha \sqrt{1-\xi^2-\eta^2} -2\Gamma(t)].
\label{spineqreg}
\end{eqnarray}
We have
\begin{eqnarray}
\Gamma(t) &=& \dot{p}q-\dot{q}p, \nonumber \\
A(t)&=& 2\frac{\dot{q}+p\Gamma(t)}{\sqrt{1-p^2-q^2}}, \nonumber \\
B(t)&=& 2\frac{\dot{p}-q\Gamma(t)}{\sqrt{1-p^2-q^2}},
\end{eqnarray}
with $\zeta=q+\imath p = \sin(\frac{i}{2})\exp(\imath \Omega)$. Here $i$ is the inclination of the orbit of a planet with respect to
the invariable plane of the planetary system and $\Omega$ is its longitude of the ascending node from an arbitrary reference direction.
We also have the precession constant of the planet's equator
\begin{equation}
\alpha = \frac{3n^2}{2\nu}J_2\Bigl(\frac{m_pR_p^2}{\mathcal{C}}\Bigr)(1-e^2)^{-3/2}.
\end{equation}
Here $J_2$ is the planet's quadrupole moment. The value of $\alpha$ depends on the quadrupole moment $J_2$ and the rotation rate.
However, $J_2$ also depends on the rotation rate (e.g. Atobe \& Ida, 2007) so that $\alpha \propto \nu$ and thus planets with
longer spin periods have longer precession periods.\\
On Earth, the three forcing periods of the Milankovi\'{c} cycles are 100~kyr, 41~kyr and 23~kyr, caused by the superposition of the
Earth's precession and the eigenfrequencies of the Solar System. We shall briefly discuss the origin of each of these frequencies and
apply this knowledge to predict what these frequencies should be for planet g.\\
To lowest order, the variations in the Earth's eccentricity and longitude of perihelion, $\varpi$, can be Fourier decomposed as (e.g.
Laskar, 1988)
\begin{equation}
 e\exp(\imath \varpi) = \sum_{n=1}^{N} M_{n,3}\exp[\imath(g_nt+\beta_n)],
 \label{edecomp}
\end{equation}
where $g_n$ are the eccentricity eigenfrequencies of the Solar System (Brouwer \& Van Woerkom, 1950), $M_{n,3}$ are the forcing
amplitudes in the Earth and $\beta_n$ are the corresponding phases.\\
A similar decomposition can be applied for the inclination and longitude of the ascending node
\begin{equation}
 \sin(\frac{i}{2})\exp(\imath \Omega) = \sum_{n=1}^{N} N_{n,3}\exp[\imath(s_nt+\delta_n)],
\end{equation}
where $s_n$ are the inclination eigenfrequencies with forcing amplitudes $N_{n,3}$ and phases $\delta_n$. For the Earth,
the largest eccentricity terms are in order $M_{5,3}$, $M_{2,3}$ and $M_{4,3}$ i.e. the terms associated with perturbations from
Jupiter, Venus and Mars (Laskar, 1988). Their frequencies are $g_5 = 4.25$~$\arcsec$/yr, $g_2 = 7.45$~$\arcsec$/yr and $g_4
= 17.78$~$\arcsec$/yr (Laskar, 1988). The superposition of these terms results in the eccentricity oscillating with three distinct
frequencies which are \ $\vert g_5-g_2 \vert$ and $\vert g_4-g_5\vert$ and $\vert g_4-g_2\vert$; the corresponding periods are
approximately 400~kyr, 128~kyr and 95~kyr respectively. The last two combine to form the 100~kyr eccentricity contribution (e.g.
McGehee \& Lehman, 2012). For the inclination, the dominant forcing is caused by Earth's eigenfrequency ($s_3=-18.85$~$\arcsec$/yr),
Mercury ($s_1 = -5.57$~$\arcsec$/yr) and Mars ($s_4 = -17.87$~$\arcsec$/yr). This leads to periods of 1.2~Myr, 106~kyr and 98~kyr.\\
We can also apply the same principle to the variations in the obliquity and precession angle, which becomes (Laskar et al., 1993)
\begin{equation}
 \sin \varepsilon\,\exp(\imath\psi) = \sum_{n=1}^{N} P_{n} \exp[\imath(\kappa_nt+\varsigma_n)].
\end{equation}
From this decomposition it turns out that $\kappa_1$ is the free precession frequency of the Earth's equator and is equal to
$\dot{\psi}_{\oplus}=-50.29$~$\arcsec$/yr (Laskar et al., 1993), with corresponding period 25.8~kyr. Now $P_{1} \gg P_{n>1}$, where
$P_{1} \sim 0.399$ corresponds to the Earth's mean obliquity of 23.5$^\circ$. The second term has a forcing frequency $\kappa_2 = s_3 =
-18.85$~$\arcsec$/yr, the inclination eigenfrequency of the Earth. Thus the obliquity will predominantly oscillate about a mean of
23.5$^\circ$ with a frequency $\vert \dot{\psi}_{\oplus}-s_3 \vert$, whose period is 41.2~kyr.\\
The last quantity in the Milankovi\'{c} cycles is the precession. The Earth's equator regresses about a fixed reference frame at a rate
of approximately 50~$\arcsec$/yr. Summer solstice occurs when the declination of the Sun, $\delta_{\odot}$, is highest. In a fixed
reference frame centred on the Earth this occurs when $\upsilon+\varpi+\psi = \pi/2$, where $\upsilon$ is the Sun's true anomaly. The
distance of the Sun to the Earth at summer solstice is then
\begin{equation}
r_{\rm SS} = \frac{a(1-e^2)}{1+e\sin(\varpi+\psi)}.
\end{equation}
Thus the precession forcing of the insolation is caused by the term $1+e\sin(\varpi+\psi)$, which contains the frequencies $\vert
\dot{\psi}_{\oplus}-g_5\vert$ and $\vert \dot{\psi}_{\oplus}-g_2 \vert$ with periods 23.7~kyr and 22.5~kyr. These two combine to form
the 23~kyr precession frequency mentioned by Milankovi\'{c} (1941).\\
On Earth the ice ages are probably triggered by changes in the summer insolation at high latitudes on the northern hemisphere (Imbrie,
1982). The typical latitude where this is computed is 65$^\circ$ N, just south of the north polar circle. Investigating the
summer insolation at high latitudes makes sense for planets with a low ($\varepsilon \lesssim 30^\circ$) obliquity because there is a
steep gradient in insolation with latitude and the ice line is situated at high latitudes. However, for planets with moderate to high
obliquity the question arises at which latitude to compute the insolation. A planet with an obliquity of 50$^\circ$ has its polar
circles at 40$^\circ$ N and 40$^\circ$ S and its equivalents of the tropics of Cancer and Capricorn are at 50$^\circ$ N and 50$^\circ$
S, meaning the declination of the star travels 100$^\circ$ over the surface of the planet between winter and summer. In addition,
planets with high ($\varepsilon \gtrsim 40^\circ$) obliquity have their annual polar insolation close to or exceeding that at the
equator. In those cases computing the summer solstice insolation at high latitude makes little sense. An excellent example is given by
Mars. Its obliquity oscillates between 15$^\circ$ and 35$^\circ$ (e.g. Ward, 1974), and its polar ice caps disappear and reappear
during periods of high and low obliquity respectively (Schorghofer, 2007). The choice of latitude can be avoided by only considering
the insolation at the pole and using the orbit-averaged value rather than the summer solstice value.\\
The yearly polar insolation is given by (Ward, 1974)
\begin{equation}
 \langle I_{\rm p} \rangle = \frac{S_*}{\pi} (1-e^2)^{-1/2}\sin \varepsilon.
\label{Q}
\end{equation}
where $S_*$ is the stellar flux at the planet's orbit i.e. $S_* = 1361(L_*/L_\odot)(1\,{\rm AU}/a)^2$~W~m$^{-2}$, $L_*$ and $L_\odot$
are the luminosities of the star and the Sun, $\varepsilon$ is the obliquity. Note that equation (\ref{Q}) has no dependence on the
precession angle of the planet, $\psi$, nor the longitude of periastron, $\varpi$, because of Kepler's second law: the decreased
insolation at apastron is balanced by the longer duration.\\
We want to introduce a few additional quantities. The insolation on planet g determines the overall temperature. Currently we
do not know if it has an atmosphere or ocean that is able to transport heat from warmer to colder areas and thus we only study the
long-term evolution of the insolation and the black-body equilibrium temperature. The instantaneous black-body equilibrium temperature
is given by (e.g. Kaltenegger \& Sasselov, 2011)
\begin{equation}
 T_{\rm eq} = T_*\Bigl(\frac{1-\mathcal{A}}{4\beta}\Bigr)^{1/4}\Bigl(\frac{R_*}{r}\Bigr)^{1/2},
\end{equation}
where $T_*$ the star's temperature, $\mathcal{A}$ is the planet's Bond albedo, $R_*$ is the star's radius, $r$ is the planet's
instantaneous distance to the star and $\beta$ represents the fraction of the planet's surface that re-radiates the absorbed flux. For
a fast-spinning planet $\beta \sim 1$ while for a tidally locked planet $\beta \sim \frac{1}{2}$. For a planet to remain habitable the
black-body equilibrium temperature needs to remain below 270~K (Selsis et al., 2007) to avoid the runaway greenhouse stage. The
orbital average is
\begin{eqnarray}
 \langle T_{\rm eq} \rangle &=&
\frac{2}{\pi}\Bigl(\frac{1-\mathcal{A}}{\beta}\Bigr)^{1/4}\Bigl(\frac{R_*}{a}\Bigr)^{1/2}\sqrt{1+e}\,
\mathcal{K}\Bigl(\sqrt{\frac{2e}{1+e}}\Bigr) \\
&\approx&
(\textstyle{\frac{1-\mathcal{A}}{\beta}})^{1/4}(\textstyle{\frac{R_*}{a}})^{1/2}[1+\textstyle{\frac{1}{2}}(\textstyle{\frac{1}{2}}e)^{
1/2 } +\textstyle{\frac{25}{16}}(\textstyle{\frac{1}{2}}e)] \nonumber \\
&+& O(e^{3/2}), \nonumber
\end{eqnarray}
where $\mathcal{K}(k)$ is the complete elliptic integral of the first kind. Note that the equilibrium temperature increases with
increasing eccentricity at fixed semi-major axis, just like the annually-averaged insolation, but it does so quicker. It is not clear
whether for habitability the instantaneous black-body equilibrium temperature needs to remain below 270~K or the orbitally-averaged
one, so we adopt the latter. These relations suggest that the extent of the habitable zone is not only semi-major axis but also also
eccentricity dependent. This appears to agree with Williams \& Pollard (2002), who suggested that the fundamental quantity for
habitability is annually-averaged insolation, and not instantaneous insolation.\\
The equilibrium temperature of the planet is different from the temperature at the pole that one may derive from the polar insolation,
whose annually-averaged value is given by (Schorghofer,2008)
\begin{eqnarray}
 \langle T_p \rangle &\approx& \frac{1}{2\sqrt{\pi}}\Bigl(\frac{S_*(1-\mathcal{A})\sin \varepsilon}{\xi
\sigma_{\rm SB}}\Bigr)^{1/4}(1-e^2)\frac{\Gamma(\frac{5}{8})}{\Gamma(\frac{9}{8})}
\nonumber \\
&\times&\Bigr[1-\frac{12}{5}\Bigl(\frac{\Gamma(\frac{9}{8})}{\Gamma(\frac{5}{8})}\Bigr)^2e\sin (\varpi+\psi)\Bigr]+O(e^2),
\label{tave}
\end{eqnarray}
which has a clear dependence on the longitude of periastron and precession angle of the planet. Here $\xi$ is the infra-red
emissivity and $\sigma_{\rm SB}$ is the Stefan-Boltzmann constant. For Earth $(1-\mathcal{A})/\xi \sim 1$, which we shall employ
later.\\ 
In this paper we are interested in planet g and not Earth and below we give an estimate of the periods of the long-term
insolation variations on planet g. We can predict what the effect of orbital forcing is on the insolation of planet g. We have
$\dot{\psi} = -\alpha \cos \varepsilon$ and so
\begin{eqnarray}
 \dot{\psi} &=& -19.2 \Bigl(\frac{M_*}{M_\odot}\Bigr)\Bigl(\frac{1\,{\rm AU}}{a}\Bigr)^3\Bigl(\frac{24\,{\rm
h}}{P_{\rm r}}\Bigr)\Bigl(\frac{m_p}{m_\oplus}\Bigr)^{-0.178} \nonumber \\
&\times&\Bigl(\frac{0.3}{\mathcal{C}}\Bigr)\frac{\cos
\varepsilon}{(1-e^2)^{3/2}} \quad \arcsec~{\rm yr}^{-1},
\end{eqnarray}
where we used $J_2 \propto \frac{R_p^3\nu^2}{Gm_p}$ (Atobe \& Ida, 2007; Murray \& Dermott, 1999), scaled it from Earth's values,
and $P_{\rm r}$ is the rotation period. This above relation is valid for rotation periods shorter than 13 days. For longer rotation
periods the distortion of the planet is no longer only dependent on the rotation and other assumptions will have to be made about
$J_2$. The value of $\dot{\psi}_{\rm g}$ is uncertain by a factor of two because of uncertainties in the mass-radius relation, moment
of inertia, mass of the planet and star.\\
Another source of uncertainty in $\dot{\psi}_{\rm g}$ is the rotation period of planet g. At low eccentricity tidal interaction
with the central star should decrease the rotation period on a specific time scale, which when $\nu \gg n$ and $\nu \gg
\gamma$ is roughly given by (Ferraz-Mello, 2013)
\begin{equation}
 \tau_{\rm gr} = \frac{16}{45}\Bigl(\frac{\nu}{n}\Bigr)^2\Bigl(\frac{m_{\rm g}}{M_*}\Bigr)\Bigl(\frac{a}{R_{\rm
g}}\Bigr)^3\Bigl(\frac{\gamma}{1\,\rm{rad}~\rm{yr}^{-1}}\Bigr)^{-1}.
\end{equation}
When treating planet g as having internal properties similar to Earth { and setting the initial rotation period equal to 1~d we
have} $\tau_{\rm gr} \sim 10$~Gyr and thus after 4.5~Gyr the rotation period is less than double the primordial value, but only if the
primordial rotation period is shorter than two days. For longer primordial rotation periods the planet will be despun in a shorter time
than the age of the system. Unless it gets trapped in a spin-orbit resonance its rotation rate will be given by the equilibrium
rotation (Ferraz-Mello, 2013)
\begin{equation}
 \frac{\nu_{\rm lim}}{n} \approx 1 + 6\frac{\gamma^2}{n^2}\Bigl(1+\frac{\gamma^2}{n^2}\Bigr)^{-1}e^2,
\end{equation}
which works out to be $\nu_{\rm lim,g} \sim 1.26n$ for $e_{\rm g} \sim 0.22$. When treating planet g as a more solid body such as Mars,
the rotation period will have barely changed either. If the planet formed through giant impacts, such as the terrestrial planets of the
Solar System (e.g. Kokubo \& Ida, 1998), we expect planet g's rotation to be rapid (Kokubo \& Ida, 2007) and to have remained so. All
this assumes that there is no large satellite, which modifies the precession and rotation considerably (Brasser et al., 2013).\\
Assuming for now that planet g's rotation period is comparable to Earth we have $J_2 \sim 0.71J_{2\,\oplus}$ and taking $\mathcal{C}
= 0.3$, intermediate between Earth and the giant planets, then $\alpha \sim 50$~$\arcsec$/yr. The values of $g_6$ and $s_6$ can be
read from Table~\ref{eigenval}. Examining the corresponding eigenvectors in Table~\ref{eigenvec} and because
$M_{i,j}\propto S_{i,j}$, one sees that $M_{6,6} \gg M_{n \neq 6,6}$, so that the eccentricity of planet g remains constant.
Hence the obliquity should oscillate with a frequency $\vert \dot{\psi}_{\rm g}-s_6\vert$ where $\dot{\psi}_{\rm g}=\alpha_{\rm g} \cos
\varepsilon$, and the precession effect should have a frequency of $\vert \dot{\psi}_{\rm g}-g_6\vert$. At low obliquity the periods
are roughly 52.2~kyr for the obliquity and 19.4~kyr for precession. These values are likely uncertain by factors of a few but a
change in the periodicities does not change the overall outcome. We shall use these expressions in Section 5.\\
After integrating the motion of the planets and converging to a planar stable solution, the third part of our method concerns the
calculation of the annual insolation at the poles. The annual polar insolation depends on the obliquity and eccentricity of the planet;
the obliquity forcing depends on the inclination. We only have a planar orbital solution for this system because the relative
inclinations are unknown. Thus to compute the obliquity variations we must make an assumption about the inclination distribution. Since
most super Earth systems appear to have mutual inclinations less than 5$^\circ$ (Tremaine \& Dong, 2012; Figuera et al., 2012), we
assume the mutual inclinations of the planets follow a Rayleigh distribution (Fang \& Margot, 2012) with a width of $\sigma_{\rm i} =
2^\circ$. This width corresponds to the current typical mutual inclinations of the planets in the Solar System (Fabrycky et al., 2012)
and appears to reproduce the observed inclination distribution of the Kepler super Earth systems (Fang \& Margot, 2012). We took
the orbital elements of the most stable planar solution and imposed an inclination on each planet. We set the longitude of the nodes
to be randomly uniform from 0 to 360$^\circ$ but kept the mean longitudes and longitudes of periastron constant. We generated 100
such configurations and ran them for 1~Myr.\\
The rotation rate of planet g is mostly unknown, so we assumed a rotation period between 6~h and 240~h. The former is consistent with
its expected primordial rotation rate if it formed by giant impacts (Kokubo \& Ida, 2007; Kokubo \& Genda, 2010) while the latter is
an imposed upper limit. The value of $J_2$ was scaled from that of Earth. When $J_2$ reaches the value for Venus (4.56$\times
10^{-6}$), which occurs when the rotation period is approximately 13~days, we do not lower $J_2$ beyond this value.\\
Equations (\ref{spineqreg}) contain no singularities and were integrated numerically using the Bulirsch-Stoer method (Bulirsch \&
Stoer, 1966) using a pre-generated planetary evolution. We refer to Brasser \& Walsh (2011) for further details.\\
The annual polar insolation and temperature are computed using the obliquity, precession angle, eccentricity and longitude of
periastron. The variation in the insolation on long time scales is computed for various initial obliquities and rotation periods.

\section{Dynamical stability analysis}
\label{ressec}
In this section we present our results from numerical simulations. We first discuss the stability analysis, then orbital influence on
the climate.\\ 
The innermost five planets in the HD 40307 system are closely packed together, with two of them (planets b and e) on
fairly eccentric orbits. Planets e and f are in a 3:2 mean motion resonance (Tuomi et al., 2013) thus the stability of the whole system
will depend sensitively on whether or not these two planets are actually in resonance. Tuomi et al. (2013) found the system is
stable when planets e and f are in a 3:2 resonance and when assuming all the planets are on circular orbits. They determined how the
stability of the system depended on the individual initial eccentricity and mean anomaly of each planet but they did not study how the
system could be made stable when taking all eccentricities and phases into account. We shall do this below.\\ 
When performing a simulation of the nominal system we find that the first two planets to have an encounter are planets e and f and
thus they are the bottleneck in any stability study. Hence we shall first investigate the stability of planets e and f.\\ 
Laskar (1997) identified that the degree of chaos in a system is directly related to the total angular momentum deficit. Thus,
after having analysed planets e and f we proceed to determine how the stability of the system depends on the initial elements of each
planet in order of decreasing specific angular momentum deficit. In other words we proceed from highest eccentricity to lowest
so that with each step we analyse the then-greatest possible source of chaos/instability.\\
The following analysis is based on more than 50\,000 numerical simulations of the system with different initial conditions. In what
follows we aim to find the most stable solution for the subsequent habitability analysis. In those cases where there are large regions
of stability we try to keep the solution that yields the lowest $\chi^2$.

\subsection{Planets e and f}
\label{ef}
\begin{figure}
\resizebox{\hsize}{!}{\includegraphics[]{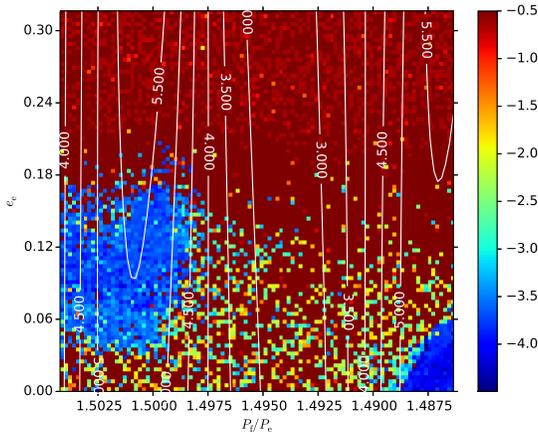}}
\caption{Image map of the maximum diffusion index, $D_{\rm max}$. The plot depicts $D_{\rm max}$ as a function of the initial period
ratio $P_{\rm f}/P_{\rm e}$ (horizontal axis) and the initial eccentricity $e_{\rm e}$ (vertical axis). The scale for the colour bar on
the right displays the values of $D$. The contours mark constant values of $\chi^2$ of the goodness of fit between the orbital solution
and the radial velocity data.}
\label{eae}
\end{figure}
We determined how the system's stability depends on the initial semi-major axis and eccentricity of planet e. The result of
this first set of simulations is depicted in Fig.~\ref{eae}. The figure displays the value of $D_{\rm max}$ as a function of initial
period ratio with planet f and initial eccentricity of planet e. The graininess of the figure is a result of the sampling resolution
and the depth of planetary encounters. High values of $D_{\rm max}$ are mostly the result of one strong encounter rather than many weak
ones. Large differences in $D_{\rm max}$ from one pixel to the next becomes a probabilistic argument of whether or not a pair of
planets underwent a deep encounter just beyond the Hill sphere during the simulation time. Longer simulations and/or a finer grid could
improve smoothness at the expense of much a longer total simulation time to process a single map. This paper is a proof of concept so
we leave these additional simulations for a future study.\\
We also want to stress the following. We approximate the orbital frequencies with the average mean motions {via Kepler's third
law}. We attempt to measure the stability directly by having the planets undergo encounters. For all of these reasons our stability
maps are grainier than those presented in Correia et al. (2005).\\
Returning to Fig.~\ref{eae}, there is only one narrow meta-stable region which is more than 1$\sigma$ away from the mean. It is at
$P_{\rm f}/P_{\rm e} < 1.489$ and $e_{\rm e} \lesssim 0.06$. The lighter blue region around $P_{\rm f}/P_{\rm e} \sim 1.5$ and $e_{\rm
e} \sim 0.1$ is unstable on longer time scales and when performing further analysis discussed below. Thus it appears that planets e and
f are {\it not} in a 3:2 mean motion resonance. {This claim is in contradiction to that of Tuomi et al. (2013) who stated that the
system was only stable if planets e and f are in resonance, and thus this begs further discussion. We can think of two explanations.\\
First, our initial conditions vary from theirs. Tuomi et al. (2013) did not perform a $\chi^2$ fit to the radial velocity data which
we calculated earlier. As a result, our initial longitudes of the planets are different from theirs, and thus the initial resonant
angles between planets e and f differ by the same amount. However, when we perform our analysis with the same initial conditions as the
nominal solution of Tuomi et al. (2013) we arrive at the same conclusion: the planets e and f are not in resonance. Second, Tuomi et
al. (2013) varied the mean motion and initial eccentricity of a single planet while keeping the eccentricities of the others fixed at
zero. Varying the mean motion causes variations in the resonant angle and larger changes in $\chi^2$. With this method it is possible
to find a solution where the resonant angle between planets e and f librates. With our initial conditions we were unable to find such
a solution that was long-term stable.\\}

\begin{figure}
\resizebox{\hsize}{!}{\includegraphics[]{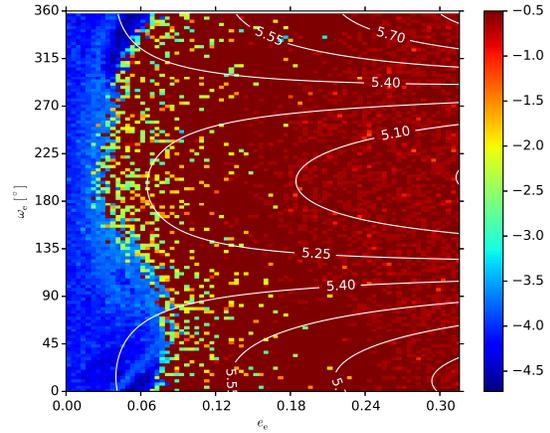}}
\caption{The same as Fig.~\ref{eae} but this time we varied the eccentricity and argument of periastron of planet e.}
\label{eeo}
\end{figure}
We want to investigate further how the stability of the system depends on the remaining parameters. To do so we update the
initial period of planet e to be equal to the most stable configuration in the lower-right region of Fig.~\ref{eae}. This requires
increasing $P_{\rm e}$ from 34.62~d to 34.80~d. This is approximately a 2.6$\sigma$ increase {when using the measured standard
deviation of the period presented in Table~\ref{orbits}}. For this stable solution the stability index $D_{\rm max}=-4.5$ and $\chi^2 =
5.36$. {The increase in $\chi^2$ from the nominal solution is caused by the longer orbital period of planet e.}\\ 
The stable island in the bottom-right corner of Fig.~\ref{eae} only exists when $e_{\rm e}$ at lower than its nominal value. Thus
the question arises whether a larger stable region, or a more stable configuration, can be found by varying $e_{\rm e}$ and
$\omega_{\rm e}$, and whether we could simultaneously also lower $\chi^2$. Thus, we then ran another set of simulations varying the
initial eccentricity and argument of periastron of planet e but using the longer orbital period of 34.80~d. The corresponding
stability map is depicted in Fig.~\ref{eeo}: there is a large region of stable motion when $e_{\rm e} <0.05$ for all values of
$\omega_{\rm e}$. From Fig.~\ref{eeo} we found that the most stable solution is at $e_{\rm e}=0.033$ and $\omega_{\rm e}=312^\circ$.
The lowering of the eccentricity is a 1.3$\sigma$ decrease with $\chi^2$ virtually unchanged {from the value obtained from
Fig.~\ref{eae} because $\chi^2$ depends much more sensitively on the period than on the eccentricity.\\
Here we want to comment on the metastable region in the left part of Fig.~\ref{eae}. When choosing the most stable configuration in the
large blue-shaded area and then performing the stability analysis varying $e_{\rm e}$ and $\omega_{\rm e}$, we were unable to find a
stable region. We performed several additional simulations with somewhat different initial conditions, but our conclusions remained the
same. For this reason we are doubtful that planets e and f could be in resonance.}\\

\subsection{Planets g and b}
\begin{figure}
\resizebox{\hsize}{!}{\includegraphics[]{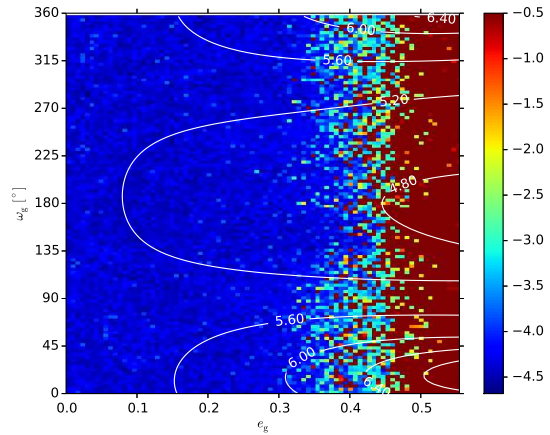}}
\caption{The same as Fig.~\ref{eae} but this time we varied the eccentricity and argument of periastron of planet g.}
\label{geo}
\end{figure}
\begin{figure}
\resizebox{\hsize}{!}{\includegraphics[]{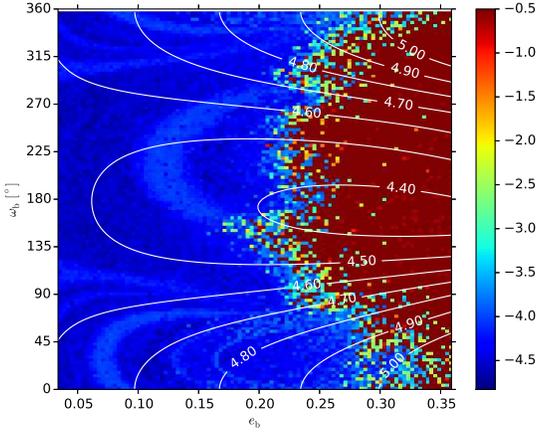}}
\caption{The same as Fig.~\ref{eae} but this time we varied the eccentricity and argument of periastron of planet b.}
\label{beo}
\end{figure}

Now that we have found a stable configuration for the planets e and f the planet with the greatest specific angular momentum is
planet g. Tuomi et al. (2013) showed that the stability of the system appears not to depend on the orbit of planet g. We confirm their
result. As long as $e_{\rm g}< 0.4$ the whole system is stable with no dependence on $\omega_{\rm g}$ (see Fig.~\ref{geo}). We only
increased $\omega_{\rm g}$ to $183^\circ$ which coincides with the lowest $\chi^2$, but left the eccentricity unchanged. We now turn to
planet b.\\
Planets c and d have fairly low eccentricities but planet b does not: its high eccentricity is most likely forced and it is likely in
apsidal alignment with planet c. To test this hypothesis we performed additional simulations where we changed the eccentricity and
argument of periastron of planet b. The result is depicted in Fig.~\ref{beo}. There is a very large region of stability for $e_{\rm b}
\lesssim 0.2$, with the most stable region occurring when $\omega_{\rm b} \sim 260^\circ$. In this region $\Delta \varpi_{\rm bc} =
\varpi_{\rm b} - \varpi_{\rm c}$ librates around 0, corresponding to apsidal alignment between planets b and c (but with a non-zero
libration amplitude). Thus, it seems stability is enhanced when one eccentricity mode is damped.\\
In Fig.~\ref{beo} the diffusion index shows little variation over a very large region in $\omega_{\rm b}$ and $e_{\rm b}$ but increases
quickly outside it. At low eccentricity there are some lighter blue regions with oxbow shapes. This alternation between more and less
stable configurations as a function of $\omega_{\rm b}$ is most likely the result of the nearby 9:4 resonance with planet c (Papaloizou
\& Trequem, 2010). We focused on those systems for which $D_{\rm max} < -4.5$ and for which $M_{1,1} \sim 0$ i.e. the eigenmode
corresponding to planet b was almost completely damped by tidal forces. We checked the amplitude of the $M_{1,1}$ mode through Fourier
analysis using the method of \v{S}idlichovsk\'{y} \& Nesvorn\'{y} (1996) and recorded those cases for which $M_{1,1} \sim 0$. From
these criteria we picked the solution with $D_{\rm max} \sim -4.5$ which has $\omega_{\rm b} = 228^\circ$ and $e_{\rm b} = 0.1541$. The
eccentricity is approximately 0.8$\sigma$ lower than the nominal value.\\

\subsection{Planets c and d}
\label{cdg}
\begin{figure}
\resizebox{\hsize}{!}{\includegraphics[]{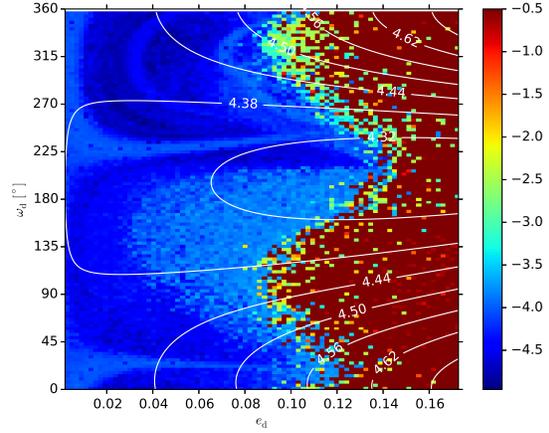}}
\caption{The same as Fig.~\ref{eae} but this time we varied the eccentricity and argument of periastron of planet d. The current
orbital elements of planet d place it in the deep blue region.}
\label{deo}
\end{figure}

We have only analysed the system's stability by changing the orbits of planets b, e and g. Changing the semi-major axis,
eccentricity and argument of periastron of planet e is mirrored by planet f because of their proximity to resonance. A similar argument
applies to planet c: it is most likely locked in apsidal alignment with planet b, and thus changing the eccentricity and argument of
periastron of planet c is mirrored by planet b. Last, the stability of the system is only compromised when $e_{\rm g}>0.4$ and
shows little dependence on its orbital configuration otherwise. \\
This leaves us with planet d, which is the central and most massive member of the inner five planets. It most strongly influences the
dynamics of the other four inner planets. Thus the current orbital elements of planet d are potentially incompatible with the most
stable configuration. We have plotted the stability map for planet d in Fig.~\ref{deo}, having changed only $e_{\rm d}$ and
$\omega_{\rm d}$. The present orbital elements place it in the deep blue, stable region for which $e_{\rm d} \lesssim 0.12$ and
$\omega_{\rm d} \lesssim 20^\circ$. The most stable region, for which $D_{\rm max} \sim -5$, occurs for $e_{\rm d}\lesssim 0.05$ and
$225^\circ \lesssim \omega_{\rm d} \lesssim 360^\circ$. Finally we took the most stable solution in the oxbow shaped region with
$e_{\rm d}=0.0441$ (0.7$\sigma$) and $\omega_{\rm d} = 249^\circ$. It has $D_{\rm max}=-4.95$ and $\chi^2=4.36$. The total deviation
of all changes from the nominal elements is 3.1$\sigma$. \\
We have now converged to a stable solution for the whole system in which planets e and f are close to but not in a 3:2 resonance.
Planets b and c are in apsidal alignment because tidal interaction with the host star has damped the amplitude of eigenmode 1. In
summary, we now have a fairly complete picture of the dynamics of the HD 40307 system in the planar case. However, is the most stable
solution unique? To that end we also analysed the stability of the nominal solution presented by Tuomi et al. (2013), but the maps are
not shown because all of these had $\chi^2 \gg 10$. We concluded that the final eccentricities of planets b, e and d are all within
0.8$\sigma$ of the values reported above, but because the initial mean longitudes of both configurations are not identical, the values
of $\omega$ for each planet were different. However, given that two different initial conditions approach a similar solution is
reassuring and we are certain that we have converged to a stable solution. In the subsection below we explore the stability of the
system by imposing a small inclination on the planet's orbits as outlined in Section~\ref{theory}. 

\subsection{Long-term trends}
\label{ltt}
\begin{figure*}
\resizebox{\hsize}{!}{\includegraphics[angle=-90]{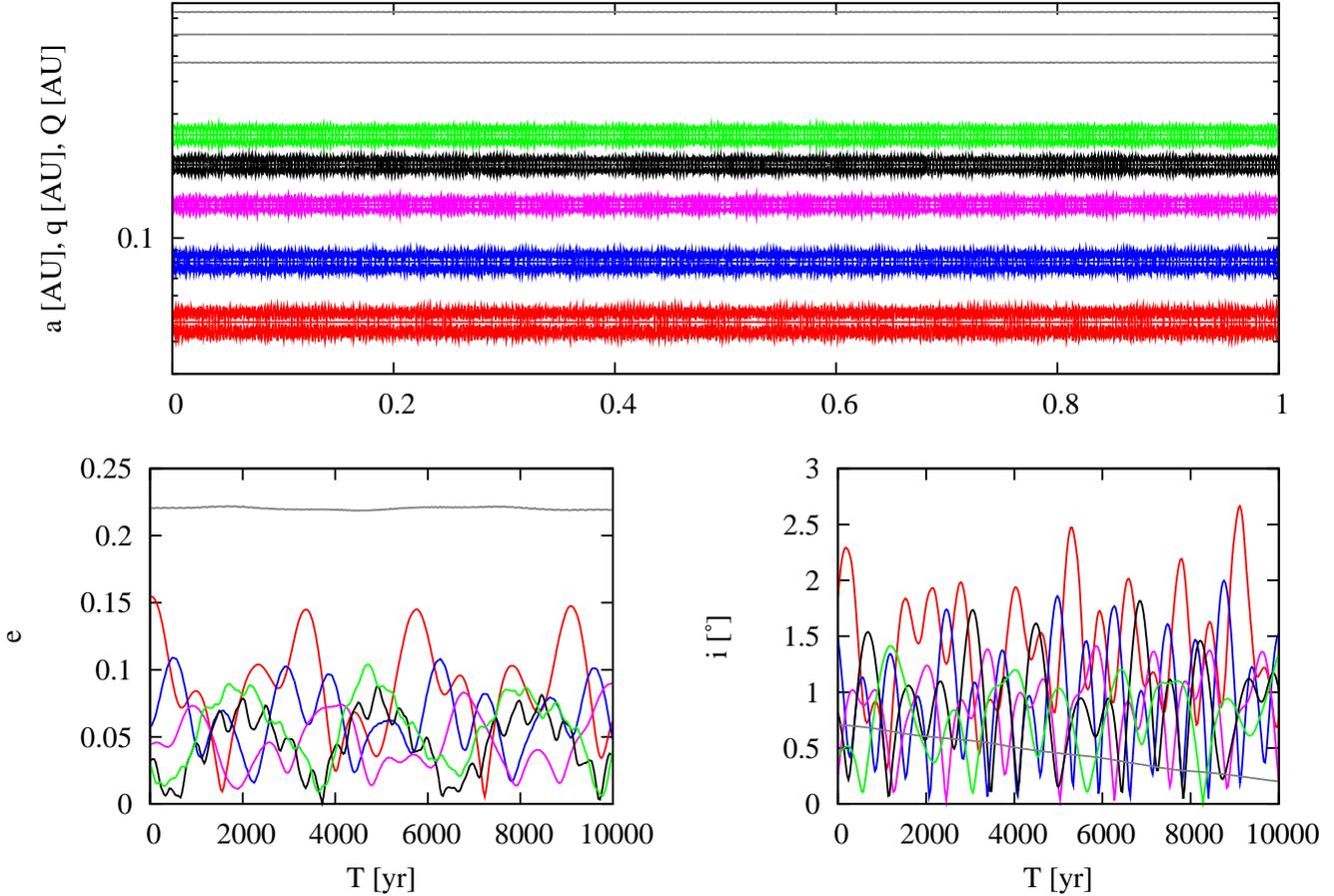}}
\caption{Evolution of the HD 40307 system for 1~Myr. The top panel shows the periastron, semi-major axis and apastron distance of all
planets. The colours are red for planet b, blue for planet c, magenta for planet d, black for planet e, green for planet f and grey
for planet g. The bottom left panel shows the eccentricity evolution for the first 10~kyr and the bottom right panel depicts the
inclinations.}
\label{evo2}
\end{figure*}

To determine long-term stability we took the most stable configuration from Fig.~\ref{deo} and integrated it for 1~Myr with inclined
orbits. The procedure to generate the initial conditions is described in Section~\ref{theory}. A long-term stable solution is
necessary to obtain the long-term insolation variation on planet g. We ran total of 100 cases and kept those that survived for the full
duration. \\
We also tested these simulations for chaos by performing frequency analysis on the eccentricity eigenfrequencies. We computed the
Fourier decomposition of the eccentricity and longitude of periastron of all planets using the method of \v{S}idlichovsk\'{y} \&
Nesvorn\'{y} (1996) over eight windows each of length 122\,880~yr. For each window we kept the frequency that had the highest amplitude
in the spectrum and computed their modified stability index $D'=\log(\sigma_g/g)$, where $g$ is the eccentricity eigenfrequency with
the strongest amplitude. A regular system should follow equation (\ref{edecomp}) and the dominant modes and amplitudes of
the Fourier decomposition should stay approximately constant and be clean delta functions (Laskar, 1988). On the other hand, for a
chaotic system the dominant frequencies will vary and their peaks will broaden. A multiplet of components can also be
viewed as a single component with varying frequency and amplitude, with the frequency of the multiplet being comparable to the
frequency spread of the multiplet. In this case the spectral lines associated with some frequencies are not clean delta functions but
exhibit a near-Gaussian profile i.e. they have significant sidebands. For a mildly chaotic system the dominant mode will show small
variations, such as in the Solar System (Laskar, 1990), but for a strongly chaotic system the dominant mode may be replaced with
another mode. This behaviour may occur when two modes have nearly equal amplitudes and the system switches between the two.\\ 
We find that 97 systems survive for 1~Myr with 18 having $D'_{\rm max} <-$3 (lowest -3.5) while the other
stable systems have $-2< D'_{\rm max} <-0.2$, suggesting that most of these configurations are chaotic, and possibly unstable on
longer time scales. The source of the chaos appears to be the alignment of planets b and c, which on long time scales switches between
circulation and libration. The planets b and c have strong interaction with planets e and f (Table~\ref{eigenvec}) and perturbations
from these planets causes the libration amplitude of $\Delta \varpi_{\rm bc}$ to vary. Sometimes the libration is broken altogether and
$\Delta \varpi_{\rm bc}$ circulates. Only at very low eccentricities of the other planets does $\Delta \varpi_{\rm bc}$ librate
continuously.\\
In the next section, we use the output from one of these simulations to compute the long-term insolation variations on planet g.
Its evolution is depicted in Fig.~\ref{evo2} and its initial conditions are listed in Table~\ref{orbits2}. The initial
eccentricities and longitudes of periastron were obtained from the stability simulations of Sections~\ref{ef} to~\ref{cdg}.
The values in brackets next to the initial eccentricities are their long-term average and standard deviation. The initial inclinations
and longitudes of the ascending node were generated by hand. Their values are indicative only. As one may see: the system is quite
stable for at least 1~Myr and most likely for much longer. { The Laplace-Lagrange AMD solution yields very similar eigenvectors for
the eccentricities but the integration constants $C^2_j$ are different, and the inclination constants $D^2_j$ are now non-zero. The
total entropy is now $\exp(Z)\sim5.5$ suggesting 5 eigenmodes out of 11 are activated (eccentricity modes 5 and 6 and inclination modes
1, 3 and 6).}\\
The last question we should answer is whether, on dynamical grounds, we can or cannot rule out the existence of another unseen planet
beyond planet g, in particular a gas giant. A rigorous dynamical test of this hypothesis is beyond the scope of the paper.
However, we performed two simple tests: one in which we placed a Saturn-mass planet on a circular orbit at 0.9~AU i.e. just 0.3~AU
away from planet g. In the second we placed a Saturn-mass planet at 1.7~AU with an eccentricity of 0.3. In the first case the
eccentricity of planet g stays approximately constant. In the second case the eccentricity of planet g oscillates between 0 and
0.22 with a period of about 20~kyr. In both cases the eccentricity of the hypothetical Saturn-mass planet remains constant and the
inner five planets remain unaffected. Thus, it appears that the system is fairly resistant to there being another, external planet. The
existence of an external, eccentric gas giant could explain planet g's eccentricity because it could be near the maximum of its
eccentricity cycle. The high eccentricity of the gas giant farther out could be the result of previous planet-planet interactions (e.g.
Juric \& Tremaine, 2008). Unfortunately, it seems unlikely to rule out the existence of another planet on dynamical grounds alone, but
observations indicate that it is unlikely such a planet exists (Tuomi et al., 2013). A planet more massive than 10~$M_{\oplus}$ up to
2~AU distance is definitely ruled out, but a 10-20~$M_{\oplus}$ object at 3~AU cannot be ruled out at this stage (Tuomi et al., 2013).
However, as stated above, the existence of such a planet or a gas giant at larger distance appears to have little effect on the
dynamics of the system and the habitability of planet g unless the nodal regression of one of these planets happens to resonate with
$\dot{\psi}_g$.
\begin{table}
\center
\caption{Initial orbital elements of the most stable dynamical solution of the HD 40307 system obtained after a limited dynamical
stability study (see Sections~\ref{ef} to~\ref{cdg}). The inclinations and longitudes of the ascending node are inserted by hand as
described in Section~\ref{theory} and are indicative only. The masses and mean anomalies have been left unchanged from
Table~\ref{orbits}. The bracketed values in the eccentricity row denote their long-term average values and standard deviations.}
\label{orbits2}
\begin{tabular}{lccc}
\hline \hline
Parameter & HD 40307 b & HD 40307 c & HD 40307 d \\
\hline
$P$ [days] & 4.3122 & 9.6183 & 20.4314\\
$e$ & 0.1541 [0.085,0.034] & 0.05778 [0.063,0.023]& 0.04408 [0.044,0.020]\\
$i$ [$^\circ$] & 1.85 & 1.49 & 0.26\\
$\Omega$ [$^\circ$] & 254.4 & 223.5 & 102.7\\
$\varpi$ [$^\circ$] & 228.0 & 234.9 & 249.0\\
$\lambda$ [$^\circ$] & 150.0 & 5.0 & 353.0\\
$a$ [AU] & 0.047524 & 0.08113 & 0.13407\\
\hline
& HD 40307 e & HD 40307 f & HD 40307 g \\
\hline
$P$ [days] & 34.8062 & 51.7768 & 196.3429 \\
$e$ & 0.03279 [0.047,0.022] & 0.03000 [0.058,0.025] & 0.2205 [0.22,0.001]\\
$i$ [$^\circ$] & 0.82 & 0.43 & 0.97\\
$\Omega$ [$^\circ$] & 220.1 & 244.6 & 187.3\\
$\varpi$ [$^\circ$] & 312.0 & 355.0 & 183.0\\
$\lambda$ [$^\circ$] & 62.0 & 23.8 & 128.0\\
$a$ [AU] & 0.1912 & 0.2491 & 0.6069\\
\hline
&$\chi^2=4.7$ \\
\hline
\hline
\end{tabular}
\end{table}

\begin{figure*}
\resizebox{\hsize}{!}{\includegraphics{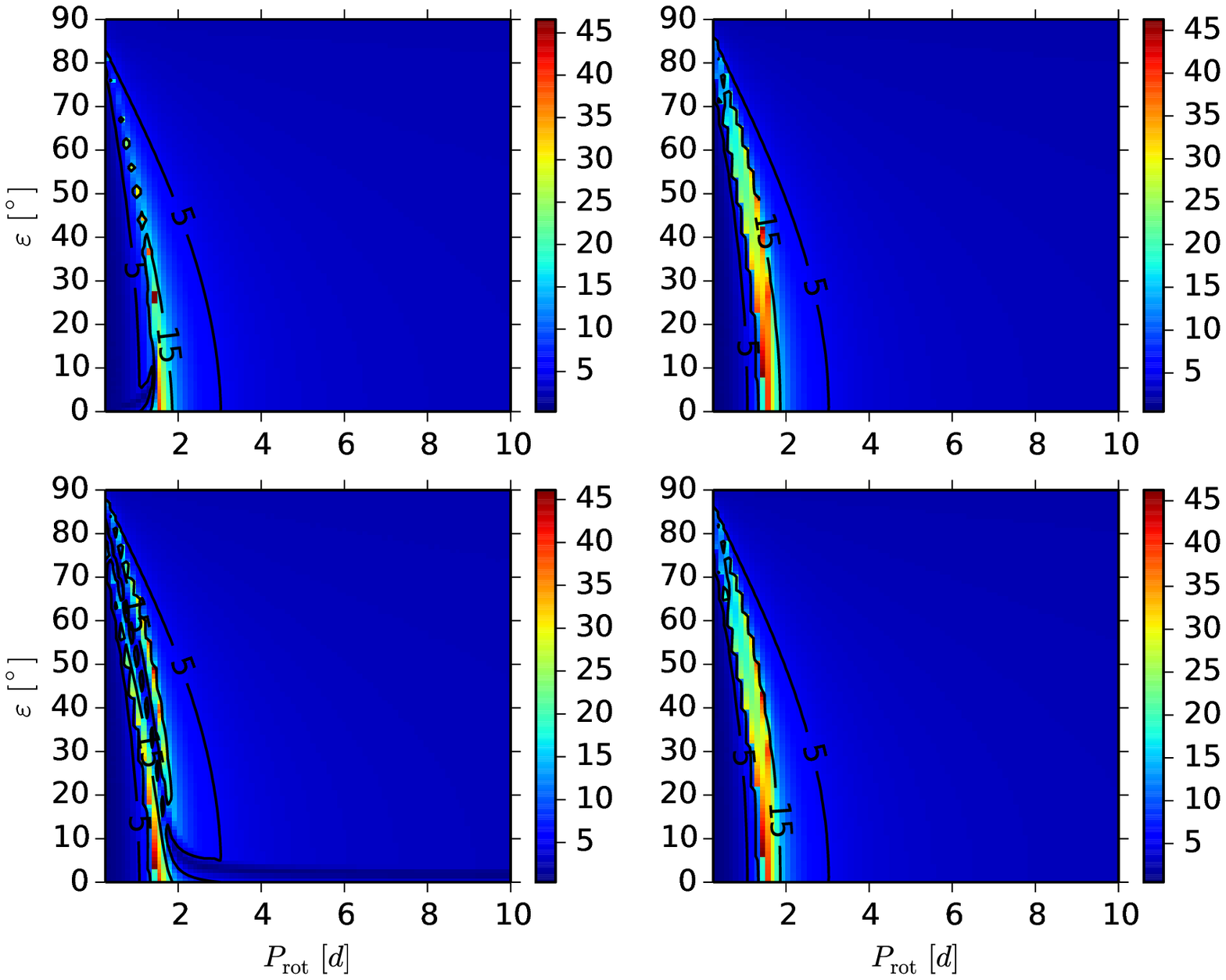}}
\caption{Contours of the amplitude of the obliquity oscillations as a function of rotation period and initial obliquity. The four
panels are for initial precession angles of 0, 90$^\circ$, 180$^\circ$ and 270$^\circ$ going clockwise from the top left. Note the two
distinct resonances when $\psi = 180^\circ$.}
\label{oamp}
\end{figure*}

\section{Obliquity and insolation evolution}
In this section we produce maps of the obliquity range, precession frequency, insolation and other quantities on planet g. The methods
were described in Sections~3 and~4: we take the output from a stable configuration that was integrated for 1~Myr and used as input to
numerically integrate the equations of the obliquity and precession evolution. We have no prior knowledge of the rotation rate of
planet g nor its $J_2$ moment, so we decided to scale these from Earth's values assuming that the fluid Love numbers are the same. We
set $J_2$ equal to Venus' value for rotation periods longer than 13~days.

\subsection{Obliquity evolution}
We display the magnitude of the obliquity oscillations as a function of rotation period and initial obliquity in Fig.~\ref{oamp}. For
rotation periods shorter than 1.5~days, longer than 2 days and low obliquities, the oscillation amplitude is generally small i.e. less
than 5$^\circ$. The oscillation period is generally of the order of 20~kyr or longer. A similar evolution is observed at long rotation
periods and high obliquities. However, the situation changes dramatically in a narrow band originating from (0.25, 85) and reaching all
the way down to (1.5,0). Here the precession of the spin pole of planet g is in resonance with the $s_6$ inclination eigenfrequency
i.e. $\dot{\psi}_{\rm g}=s_6$, and the planet is in a Cassini state (Colombo, 1966), specifically Cassini state 2. Inside Cassini state
2 the obliquity oscillates with large amplitude. For obliquities close to zero, the planet is in Cassini state 1, such as Mercury (e.g.
Correia \& Laskar, 2010). For planet g the only substantial forcing on the obliquity is caused by the $N_{6,6}$ term (which is the free
inclination $i_{\rm f}$). For Fig.~\ref{oamp} it was $i_{\rm f}=0.58^\circ$. The obliquity at which the states occur are approximately
given by (Ward \& Hamilton, 2004)

\begin{equation}
 \varepsilon \approx \arctan\Bigl(\frac{\sin i_{\rm f}}{1\pm \alpha/s_6}\Bigr), \quad \varepsilon \approx \pm \arccos \Bigl(\frac{s_6
\cos i_{\rm f}}{\alpha}\Bigr).
\end{equation}
The first two roots correspond to states 1 and 3 (state 3 is retrograde), while the second formula approximates states 2 and 4. Note
that states 1 and 4 do not exist when $\alpha/s_6 < (\sin^{2/3} i_{\rm f}+\cos^{2/3} i_{\rm f})^{3/2} \approx 1.07$,
corresponding to $\alpha = 31.5$~$\arcsec$/yr, with a matching critical obliquity $\tan \varepsilon_{\rm c} = \tan^{1/3} i_{\rm f}$
i.e. $\varepsilon_{\rm c} \approx 12.2^\circ$. For the parameters we chose states 1 and 4 disappear when the rotation period is nearly
1.5 days, which matches Fig.~\ref{oamp} above.\\

\begin{figure*}
\resizebox{\hsize}{!}{\includegraphics{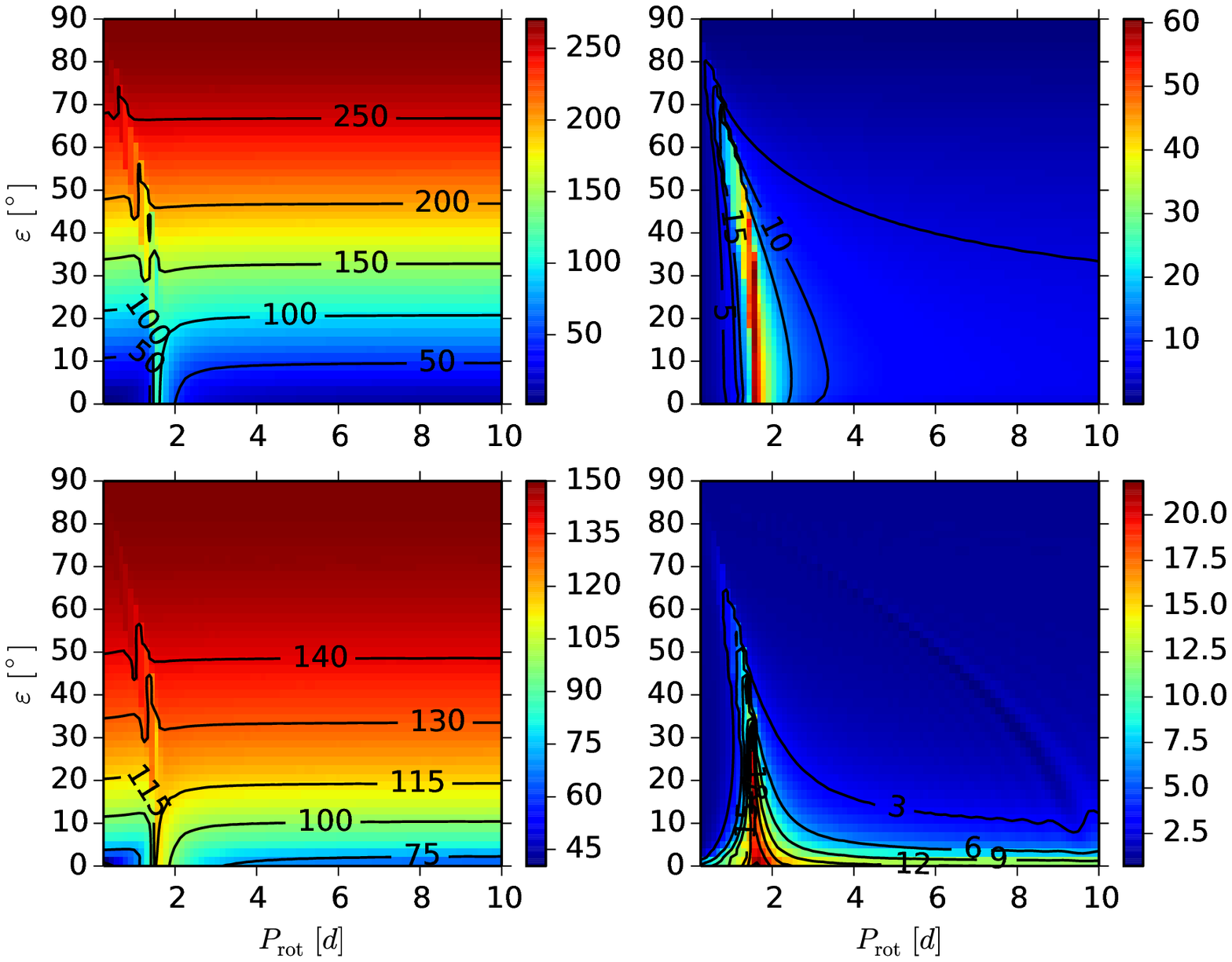}}
\caption{The variations in polar insolation (top left) and its standard deviation (top right) as a function of rotation period and
obliquity. The bottom left panel depicts the variation in polar black body temperature while the bottom right panel shows its standard
deviation.}
\label{qyamp}
\end{figure*}

\subsection{Polar insolation and black-body temperature}
How do these variations in obliquity affect the yearly insolation and temperature at the poles? We display the results in the
following figures. In the top-left panel of Fig.~\ref{qyamp} we display the average yearly insolation in Watts per square metre
at the North pole as a function of the rotation period (horizontal) and initial obliquity (vertical). The initial precession
angle was set to $\psi=90^\circ$. The scale is displayed by the vertical bar. The region where large excursions in the obliquity
occur are mirrored in this figure. The top-right panel shows contours of the standard deviation of the insolation at the North pole.
These contours follow those of the obliquity: the yearly-averaged insolation only depends on the obliquity. The bottom-left
panel displays the average black-body temperature at the North pole in Kelvin obtained from equation (\ref{tave}), while the
bottom-right shows the standard deviation of the black-body temperature range. Thus for obliquities above 30$^\circ$, the average
North pole black-body temperature is between 125~K and 150~K but the standard deviation of the long-term temperature range is about
20~K, so the expected long-term peak-to-peak changes could have amplitudes of up to 50~K. The long-term insolation
variations are $\sim$20~W unless the planet is in the Cassini state.\\

\begin{figure}
\resizebox{\hsize}{!}{\includegraphics[angle=-90]{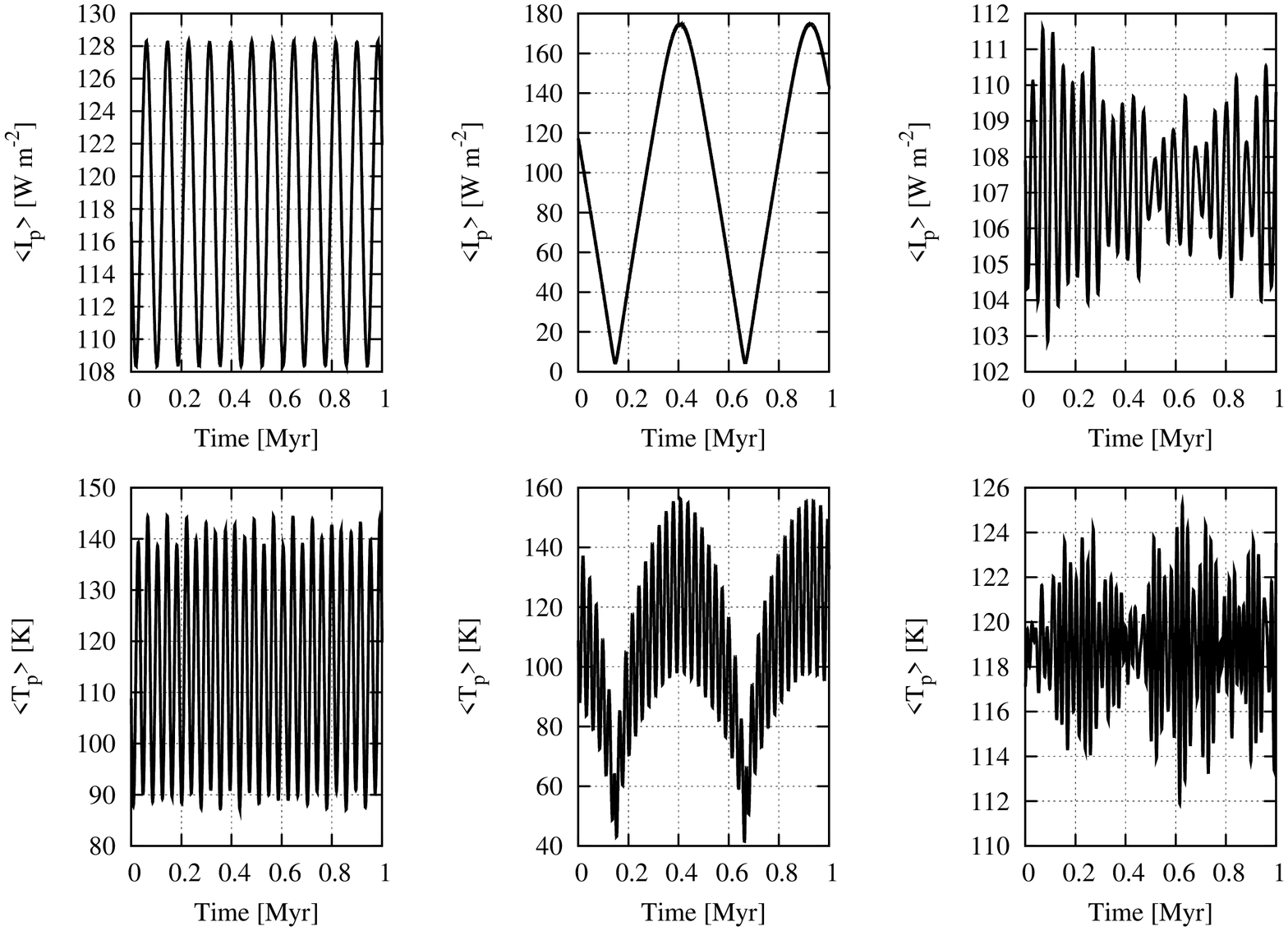}}
\caption{The evolution of the annually-averaged polar insolation of planet g for a 3~d rotation period (top left), 1.5~d rotation
period (top middle) and Earth (top right), and the polar black-body temperature (bottom left, middle and right).}
\label{insol3d}
\end{figure}
How does the insolation and black-body temperature on planet g compare to Earth's? { We created a plot of the yearly insolation at
the North pole for planet g and Earth (top row) and their black-body temperatures (bottom row). We fixed the initial obliquity of
planet g at 25$^\circ$. The rotation period was 3~d for the left column and 1.5~d for the middle column, inside the Cassini state}.\\
In the first case, the amplitude of the insolation variation is comparable to that of Earth, but the periodicities are very
different: for Earth the dominant periodicity is 41~kyr, with modulations of 39~kyr and 53~kyr, while for planet g the period is
approximately 102~kyr (frequency $\vert \dot{\psi}_{\rm g}-s_6 \vert \sim 12.7$~$\arcsec$/yr). The situation is different for the
evolution of the black-body temperature: the periodicity for planet g is 40~kyr caused by precession (frequency $\vert \dot{\psi}_{\rm
g} - g_6 \vert \sim 32.0$~$\arcsec$/yr), with the 102~kyr modulation on top. For the Earth the precession component is 23~kyr,
modulated by obliquity at 41~kyr. However, there is a striking difference between the left top and bottom panels. While for both planet
g and Earth the annual average polar insulation is similar and shows little variation, the same cannot be said for the polar black-body
temperature. For Earth this shows a 10~K variation on long time scales -- seasonal variations are not taken into account -- but for
planet g the range is 50~K! If Earth's 10~K variations can cause regular ice ages, the effect on planet g appears much more enhanced
and would likely occur on a similar time scale. The insolation variation on planet g is substantially different from Earth's
because of its high eccentricity. Indeed, the temperature deviations are mostly precession driven, while for Earth these are
driven both by precession and obliquity because the amplitude of eccentricity oscillations is not much larger than the obliquity
oscillations.\\
When the rotation period of planet g is 1.5~d and the planet is in the Cassini state, the situation is changed dramatically: its
obliquity oscillates with a large ($\sim 25^\circ$) amplitude and period 450~kyr. The polar insolation now varies by a factor of 10 and
the polar black-body temperature variations are also much larger. The short-period variations are caused by the precession, as they
were in the 3~d rotation case, but the long-periodic variation is caused by the obliquity. The largest difference with the 3~d rotation
case is the long-term variation in the polar insolation. If the climate change is primarily driven by changes in polar insolation
rather than temperature variations then the 1.5~d rotation case could cause more extreme climate variations. As we argued in
Section~\ref{theory}, during periods of high obliquity the equator becomes colder than the pole, which could cause a dramatic shift in
ice deposits from the poles to the equator (as happens on Mars; Schorghofer, 2007). In addition, the high-obliquity causes the polar
circles to be near the equator and the tropics of `Cancer' and `Capricorn' to be near the pole, causing rapid and extreme seasonal
temperature variations. Once again taking Mars as an example, its extreme obliquity variations inhibit the climate's stability because
of the melting and freezing of the polar caps, and the exchange of ice between the poles and the equator (Schorghofer, 2007). Thus we
cautiously argue against the 1.5~d case being suitable for habitability unless the planet happens to be at the exact Cassini state
equilibrium where the obliquity is constant.\\
So far we have only analysed the variation in insolation for a rotation period of 3~d and 1.5~d i.e. outside and inside the Cassini
state. There is also a domain of shorter rotation periods. For these shorter rotation periods the climate cycles occur at a
faster pace. For example, when the rotation period is just 12~h the precession periodicity approaches 10~kyr and the obliquity
variations occur on a time scale of 20~kyr. 

\section{Discussion}
In the previous sections we analysed the dynamical stability of the HD 40307 super Earth system using numerical methods. We followed
this up with a detailed dynamical study on the precession, obliquity and insolation variations on planet g. Several of our results
require further elaboration.\\
First we discuss our methods. In this study we did not vary the masses of the planets for two reasons. First, changing the masses
would have added an extra dimension to our numerical simulations and the parameter space would have pushed the limits of our
computational resources. Second, the stability of the system depends on a low power of the planetary masses (Chambers, 1996; see
Section~2). The only quantity that depends linear on the masses are the precession frequencies. The uncertainty in the masses are less
than factors of two and thus these quantities change by a similar amount. Similarly, it has been estimated that the $\sin I$
effect typically increases the masses by an amount lower than 20\% (Tremaine \& Dong, 2012), which is also not enough to cause
substantial changes in the dynamics. Thus we decided to use the nominal masses for this study.\\
The next topic pertains to the inclinations of the planets. The RV data typically are unable to constrain the mutual inclinations
of the system, unless it has reached a fixed point through tides (Batygin \& Laughin, 2011). That is likely not the case here, so we
introduced the inclinations by hand as explained in Section~\ref{theory} based on the method of Fang \& Margot (2012). They
demonstrated that their Rayleigh distribution matched well with the Kepler super Earth systems. We could have chosen to use a different
distribution which would probably have worked just as well, but this choice would have required additional justification. We caution
against increasing the maximum inclination much above $\sim 5^\circ$, because this would violate constraints that we have from planet
formation theories. It would also add a substantial amount of Angular Momentum Deficit to the system which could be exchanged with
eccentricity and ultimately make the system less stable (Laskar, 1997).\\
The possible presence of additional planets also requires some elaboration. As we said in Section~5 a Saturn-mass planet on an
eccentric orbit within 2~AU of the star is able to significantly perturb planet g. A low-eccentricity Saturn-mass planet will have
little to no effect. But what about other planets beyond 3~AU, especially gas giants just like in our Solar System? The distant spacing
of such gas giant planets would cause them to have little dynamical effect on planet g and the planets inside of it, unless one of the
eigenfrequencies of these hypothetical gas giants is commensurate with $g_6$ (or $s_6$). Such a secular resonance would increase planet
g's eccentricity to a high value and could make it cross the orbits of the inner 5 planets. This would require a coincidence between
the eigenfrequencies which is unlikely to be the case. Thus we suggest that the addition of gas giant planets beyond 3~AU does not
pose a large risk to the dynamical stability of planet g and the innermost planets.\\
The next issue that warrants discussion is the uncertainties in the parameters. The largest uncertainties are in the radius and
rotation period of planet g. Both of these directly affect the precession period. Increasing the precession period by a factor of two
implies an increase in the planetary radius by 25\%, larger than the errors presented in Sotin et al. (2007). Concerning the
rotation period, we have demonstrated in Section~2 that if planet g formed rotating quickly, it will have most likely remained so. Only
if it formed with a long rotation period could it be despun. Thus our estimates of a factor of two uncertainty in $\dot{\psi}_{\rm g}$
is probable. Increasing $\dot{\psi}_{\rm g}$ by a factor of two lowers the precession and obliquity oscillation periods by a similar
amount, reducing them from 40~kyr to 20~kyr and from 102~kyr to 51~kyr at a 3~d rotation period. All of these are comparable to Earth's
forcing frequencies and our conclusions remain the same.\\
We now turn to the habitability of planet g. Due to a lack of information about its interior and exterior we made certain assumptions,
based on recent results in the literature, about how the planet's obliquity would respond to orbital forcing. With no knowledge
about oceans or the atmosphere we can only investigate the long-term behaviour of the insolation and black-body temperature at a given
latitude, either their yearly-averaged values or those at summer solstice and compare them with Earth's. For mathematical
simplicity we took the values at the north pole. We find that the high eccentricity and quick precession of planet g because of
its proximity to the star cause substantial changes in polar black-body temperature on a time scale of up to a few tens of thousands of
years. For rotation periods of 1.5 days the obliquity displays large-amplitude oscillations with long period which also dramatically
affect the climate.\\
A third question pertains to what our results imply. Using the nominal eccentricity of planet g, and reasonable assumptions about its
$J_2$ moment, outside of the Cassini state its axial precession is comparable to Earth's and the polar black-body temperature varies on
time scales of $\sim 40$~kyr if the rotation period is a few days, and shorter time scales for a shorter rotation period. Given
that the magnitude of these black-body temperature variations are much larger than Earth's, we speculate that this causes
equally drastic but periodic changes in the climate, with the polar regions being affected much more strongly than the tropics. The
most likely consequences depend on the planet's geography and atmosphere. For a dry planet such as Mars, assuming a thin
atmosphere, the most severe effect is the reappearance and disappearance of the polar ice caps and high and low obliquity. For an
ocean planet a similar effect may occur. It has been suggested that for planet g to have a temperature similar to Earth's it
requires a 10 ~bar carbon dioxide atmosphere (Tian, 2013). Such a thick atmosphere would most likely buffet the insolation variations
induced by planetary perturbations because it has much more efficient heat transport than Earth's thin atmosphere.\\

\section{Summary and conclusions}
We investigated the dynamical stability of the HD 40307 planetary system with the aid of numerical simulations. Once a stable
solution was found it was used to determine the long-term insolation variation of planet g, which is situated in the habitable zone of
the star. We found that the most stable orbital solution of the whole system requires a 2.6$\sigma$ increase in the period of planet e.
This places planet e outside of a 3:2 mean-motion resonance with planet f. {It further requires a reduction in its eccentricity of
1.3$\sigma$.}\\ 
The high eccentricity of planet b is the result of forcing from the other planets, mostly from planet c. Its own eccentricity
eigenmode is most likely damped by tides and thus it is in apsidal alignment with planet c. {The most stable configuration of the
system requires some further reduction in the eccentricities of planets b and d. It is 3.1$\sigma$ from the nominal solution with a
reduced $\chi^2 = 4.36$.}\\
The Milankovi\'{c} cycles on planet g manifest themselves with a period similar to those on Earth, but the polar black-body temperature
variations are much more intense than on Earth because of planet g's high eccentricity. For this reason we cautiously conclude that
planet g may not be very habitable at high latitudes when the obliquity is low, thereby reducing its overall habitability. The high
eccentricity could cause regular, intense ice ages and severe ocean level changes on a wet planet such as Earth and regular
disappearance and reappearance of dry polar ice caps such as on Mars. While the periodicities are uncertain by factors of a few, the
variation in the insolation is not and thus the overall conclusion remains the same. If planet g formed with a fast rotation through a
giant impact stage, the rotation likely would have remained fast, causing rapid precession and short periods between ice ages, as well
as reducing heat transport from warmer to cooler regions (Williams, 1988).

\section{Acknowledgements}
{\footnotesize We thank Mikko Tuomi for stimulating discussions and for sharing the HARPS-TERRA radial velocity data of HD 40307.
We thank Sylvio Ferraz-Mello for discussions and the sharing of software to compare the simulations and the radial velocity data.
Last, and most importantly, we are deeply grateful to a tough and fair anonymous reviewer for his/her patience and very valuable
feedback that improved this work substantially. RB thanks SI and EK for their hospitality during several trips to Tokyo. The
Condor Software Program (HTCondor) was developed by the Condor Team at the Computer Sciences Department of the University of
Wisconsin-Madison. All rights, title, and interest in HTCondor are owned by the Condor Team.}\\

\section{Bibliography}
\footnotesize{
Abe-Ouchi A., Saito F., Kawamura K., Raymo M. E., Okuno J., Takahashi K., Blatter H., Nature 500, 193.\\
Agnor C.~B., Lin D.~N.~C., 2012, ApJ, 745, 143\\
Anglada-Escud{\'e} G., Butler R.~P., 2012, ApJS, 200, 15\\
Atobe K., Ida S., Ito T., 2004, Icar, 168, 223\\
Atobe K., Ida S., 2007, Icar 188, 1\\
Batalha N.~M., et al., 2013, ApJS, 204, 24\\
Batygin K., Laughlin G., 2011, ApJ, 730, 95\\
Batygin K., Morbidelli A., 2013, AJ, 145, 1\\
Beaug{\'e} C., Ferraz-Mello S., Michtchenko T.~A., 2012, RAA, 12, 1044\\
Borucki W.~J., et al., 2011, ApJ, 728, 117\\
Brasser R., Walsh K.~J., 2011, Icar, 213, 423\\
Brasser R., Ida S., Kokubo E., 2013, MNRAS, 110\\
Brouwer, D., van Woerkom, A. J. J. 1950. Astron. Papers Amer. Ephem. 13, 81-107.\\
Bulirsch, R., Stoer, J. 1966. \ Numerische Mathematik 8, 1-13.\\
Chambers J.~E., Wetherill G.~W., Boss A.~P., 1996, Icar, 119, 261\\
Chambers J.~E., 2001, Icar, 152, 205\\
Chiang E., Laughlin G., 2013, MNRAS, 431, 3444\\
Colombo G., 1966, AJ, 71, 891\\
Correia A.~C.~M., et al., 2005, A\&A, 440, 751\\
Correia A.~C.~M., Laskar J., 2010, Icar, 205, 338\\
Fabrycky D.~C., et al., 2012, arXiv:1202.6328\\
Fang J., Margot J.-L., 2012, ApJ, 761, 92\\
Ferraz-Mello S., 2013, CeMDA, 116, 109\\
Figueira P., et al., 2012, A\&A, 541, A139\\
Gaidos E., Fischer D.~A., Mann A.~W., L{\'e}pine S., 2012, ApJ, 746, 36\\
Hansen B.~M.~S., Murray N., 2012, ApJ, 751, 158\\
Howard A.~W., et al., 2010, Sci, 330, 653\\
Howard A.~W., et al., 2012, ApJS, 201, 15\\
Huybers, P., Wunsch, C., 2005, Nature 434, 491\\
Huybers, P., 2011, Nature 480, 229\\
Ida S., Lin D.~N.~C., 2010, ApJ, 719, 810\\
Imbrie J., Imbrie J.~Z., 1980, Sci, 207, 943\\
Imbrie J., 1982, Icar, 50, 408\\
Juri{\'c} M., Tremaine S., 2008, ApJ, 686, 603\\
Kaltenegger L., Sasselov D., 2011, ApJ, 736, L25\\
Kasting J.~F., Whitmire D.~P., Reynolds R.~T., 1993, Icar, 101, 108\\
Kley W., Nelson R.~P., 2012, ARA\&A, 50, 211\\
Kokubo E., Ida S., 1998, Icar, 131, 171\\
Kokubo E., Ida S., 2007, ApJ, 671, 2082\\
Kokubo E., Genda H., 2010, ApJ, 714, L21\\
Laskar J., 1988, A\&A, 198, 341\\
Laskar J., 1990, Icar, 88, 266\\
Laskar J., 1993, CeMDA, 56, 191\\
Laskar J., et al., 1993, A\&A, 270, 522\\
Laskar J., 1997, A\&A, 317, L75\\
Levison H.~F., Duncan M.~J., 1994, Icar, 108, 18\\
Lissauer J.~J., et al., 2011, ApJS, 197, 8\\
Lithwick Y., Wu Y., 2012, ApJ, 756, L11\\
Lopez E.~D., Fortney J.~J., Miller N., 2012, ApJ, 761, 59\\
Lovis C., et al., 2011, A\&A, 528, A112\\
McGehee, R., Lehman, C., 2012, SIAM J. Appl. Dyn. Syst. 11, 684–707\\
Marzari F., Tricarico P., Scholl H., 2003, MNRAS, 345, 1091\\
Mayor M., Queloz D., 1995, Natur, 378, 355\\
Mayor M., et al., 2003, Msngr, 114, 20\\
Mayor M., et al., 2009, A\&A, 493, 639 \\
Mayor M., et al., 2011, arXiv, arXiv:1109.2497\\
Michtchenko T.~A., Malhotra R., 2004, Icar, 168, 237\\
Milankovi\'{c}, M., 1941. Royal Serbian Sciences 33,633\\
Morbidelli A., Brasser R., Tsiganis K., Gomes R., Levison H.~F., 2009, A\&A, 507, 1041\\
Neron de Surgy, O., Laskar, J., 1997, A\&A 318, 975\\
Nobili A., Roxburgh I.~W., 1986, IAUS, 114, 105\\
Papaloizou J.~C.~B., Terquem C., 2010, MNRAS, 405, 573\\
Petrovich C., Malhotra R., Tremaine S., 2013, ApJ, 770, 24\\
Raymond S.~N., Barnes R., Mandell A.~M., 2008, MNRAS, 384, 663\\
Rein H., 2012, MNRAS, 427, L21\\
Saha P., Tremaine S., 1994, AJ, 108, 1962\\
Schorghofer N., 2007, Natur, 449, 192\\
Schorghofer N., 2008, GeoRL, 35, 18201\\
Selsis F., Kasting J.~F., Levrard B., Paillet J., Ribas I., Delfosse X., 2007, A\&A, 476, 1373\\
{\v S}idlichovsk{\'y} M., Nesvorn{\'y} D., 1996, CeMDA, 65, 137\\
Sotin C., Grasset O., Mocquet A., 2007, Icar, 191, 337\\
Tian, F., 2013, Science China, Earth Sciences, 56, 1\\
Tremaine S., Dong S., 2012, AJ 143, 94\\
Tuomi M., Anglada-Escud{\'e} G., Gerlach E., Jones H.~R.~A., Reiners A., Rivera E.~J., Vogt S.~S., Butler R.~P., 2013, A\&A, 549, A48\\
Valencia D., O'Connell R.~J., Sasselov D., 2006, Icar, 181, 545\\
Ward, W.~R., 1974, JGR, 79, 3375\\
Ward W.~R., Hamilton D.~P., 2004, AJ, 128, 2501\\
Williams D.~M., Kasting J.~F., 1997, Icar 129, 254\\
Williams D.~M., Pollard D., 2003, IJAsB, 2, 1\\
Williams G.~P., 1988, ClDy, 3, 45\\
Wisdom J., Holman M., 1991, AJ, 102, 1528\\
Wu Y., Lithwick Y., 2011, ApJ, 735, 109}
\end{document}